\documentclass[useAMS,usenatbib,usegraphicx]{mn2e} 
\usepackage{amssymb} 
\usepackage{amsmath}
\usepackage{amsfonts}
\usepackage{times} 
\usepackage{aas_macros} %needed to define journal macros generated by 
\newcommand{\cd}{d$^{-1}$}
\newcommand{\msun}{M$_\odot$}

\usepackage[usenames]{color}

\title[SuperWASP sdBV discoveries]{Three new pulsating sdB stars discovered with SuperWASP\thanks{Based on observations made with the Mercator-, Nordic Optical-, and William Herschel Telescopes all operated on the island of La Palma, Spain.}}
    
\author[D.~L.~Holdsworth et al.] 
{Daniel L. Holdsworth,$^{1,2}$\thanks{Email: dlholdsworth@uclan.ac.uk} Roy H. \O stensen$^3$, Barry Smalley$^2$ and John H. Telting$^4$ \\
$^{1}$Jeremiah Horrocks Institute, University of Central Lancashire, Preston PR1 2HE, UK \\
$^{2}$Astrophysics Group, Keele University, Staffordshire ST5 5BG, UK\\
$^{3}$Department of Physics, Astronomy, and Materials Science, Missouri State University, Springfield  MO 65804, USA\\
$^{4}$Nordic Optical Telescope, Rambla Jos{\'e} Ana Fern{\'a}ndez P{\'e}rez 7, E-38711 Bre{\~n}a Baja, Spain}

\pagerange{\pageref{firstpage}--\pageref{lastpage}} \pubyear{2017} 

\begin{document} 

\label{firstpage}
\maketitle 

\begin{abstract} 
We present an analysis of three new pulsating subdwarf B stars discovered in the SuperWASP archive. Two of the stars, J1938+5609 and J0902--0720, are p-mode pulsators; J1938+5609 shows a pulsation at 231.62\,\cd\, ($P=373$\,s; 2681\,$\umu$Hz) with an amplitude of 4\,mmag, whereas J0902--0720 pulsates at frequencies 636.74\,\cd\, ($P=136$\,s; 7370\,$\umu$Hz) and 615.34\,\cd\, ($P=140$\,s; 7122\,$\umu$Hz), with amplitudes 7.27 and 1.53\,mmag, respectively. The third star, J2344--3427, is a hybrid pulsator with a p-mode frequency at 223.16\,\cd\, ($P=387$\,s; $2583\umu$Hz) and a corresponding amplitude of 1.5\,mmag, and g\,modes in the frequency range $8.68-28.56$\,\cd\, ($P=3025-9954$\,s; $100-331\umu$Hz) and amplitudes between 0.76 and 1.17\,mmag. Spectroscopic results place J1938+5609 and J2344--3427 among the long-period or hybrid pulsators, suggesting there may be further modes in these stars below our detection limits, with J0902--0720 placed firmly amongst the p-mode pulsators. 
\end{abstract} 


%%%%%%%%%%%%%%%%%%%%%%%%%%%%%%%%%%%%%%

\section{Introduction}

The hot subdwarf B (sdB) stars are evolved low-mass ($M\lesssim0.5$\,\msun) stars that consist of helium burning cores and a thin hydrogen atmosphere which is unable to support H-shell burning \citep{heber84,heber86}. They are found on the Extreme Horizontal Branch (EHB) with temperatures between about 22\,000 and 40\,000\,K, and surface gravities between values of $\log g=5.0$ to $6.2$. The progenitors to these stars, main-sequence stars with masses $\lesssim2$\,\msun, have undergone a core helium flash and made their way to the Red Giant Branch (RGB). Here they retain approximately 1\,per\,cent of their hydrogen mass. As a result of this low hydrogen mass, after core helium exhaustion, the stars move directly to the White Dwarf (WD) cooling track, rather than moving to the Asymptotic Giant Branch phase. The lifetime for the EHB stars is short: to move from the zero-age EHB to the terminal-age EHB takes between 100 and 150\,Myr. 

The formation of sdB stars is still somewhat of a mystery, in particular the mechanism which causes only a small amount of H to remain in an envelope while the He core mass is of order $0.5$\,\msun. \cite{mengel76} suggested mass-loss in Roche lobe filling binaries could account for the resulting EHB star, while \citet{webbink84} suggested WD mergers as a possible progenitor. More recently, \citet{maxted01,maxted02} found evidence that a large number of sdB stars are found in binaries, adding weight to the binary formation theory. Following from that discovery, several mechanisms have been proposed for the resulting low-mass  H envelope of the EHB stars \citep{han02,han03}. For a thorough overview of the sdB stars as a whole, see \citet{heber09,heber16}.

There exists a subgroup of the sdB stars which show light variations on the order of seconds to hours: the subdwarf B variable (sdBV) stars. The prototype for this class, EC\,14026--2647, was observed by \citet{kilkenny97} as part of the Edinburgh-Cape (EC) survey \citep{stobie97b}. They found the star to be pulsating with a period of $144$\,s ($\nu=600$\,\cd; $6944\,\umu$Hz) and an amplitude of 12 millimagnitudes (mmag). 

As the number of sdBV stars grew, it became clear that two types of pulsators exist. The initial set of sdBV stars showed very short period pulsations which were identified as pressure-mode (p-mode) oscillations by \citet{charpinet97}. These modes are driven in the envelopes of the stars. A later discovery by \citet{green03} showed a longer period variation in the sdB star PG\,$1716+426$. The period in this, and other stars, was of the order 10 times longer than those previously seen in the sdB stars, and there was a distinct difference between the temperatures of the initial group of pulsators and this new one. These observations led to the conclusion that the newly identified sdBVs were gravity-mode (g-mode) pulsators \citep{fontaine03}. Since the first discovery of variability in the subdwarf stars, 110 sdBV, sdOBV and sdOV  stars have been identified.  Table\,\ref{tab:sdBV-cat} list the known sdBV and sdOV stars (hereafter sdV when referred to jointly).

\begin{table*}
\sevensize
 %\tiny
 %\scriptsize
    \caption{Catalogue of the sdV stars, updated from \citet{ostensen10b}. The pulsator type is given as: P\,=\,p-mode, G\,=\,g-mode and H\,=\,hybrid, targets with two types are explained in the text. The three targets in bold font are those analysed in this paper. References are either the discovery paper, or a subsequent paper with further information. Where there are multiple sources of stellar parameters, a weighted mean is given.}
\label{tab:sdBV-cat}
  \begin{tabular}{llccccccccc}
    \hline
	\multicolumn{1}{l}{Name} & \multicolumn{1}{l}{Alternative} & $m_V$ & $T_{\rm eff}$ & $\log g$               & $\log(N_{\rm He}/N_{\rm H})$ &Period range & N$_P$ & Pulsator & \multicolumn{2}{c}{Reference}  \\
		                                  & \multicolumn{1}{l}{name}       &              & (kK)               &   (cm\,s$^{-2}$)    &                                                &(s)                  &             &   Type    & Pulsation             & Parameters \\
		  \hline

V429\,And    		& HS\,0039+4302   		& 15.1 	& $32.4\pm0.7$ 	& $5.70\pm0.10$ 	& $-2.20\pm0.20$	& 134--242 	& 14	& P	& 1, 2, 3 & 1 \\
BI\,Ari			& PG\,0154+182 		& 15.3 	& $35.8\pm0.3$ 	& $5.80\pm0.10$ 	& $-1.67\pm0.10$	& 111--164 	& 6 	& P	& 4, 5 &  6 \\
KN\,Boo			& PG\,1419+081 	        	& 15.1 	& $33.3\pm0.3$ 	& $5.85\pm0.10$ 	& $-1.75\pm0.10$	& 136--143 	&  2 	& P	& 7 & 6 \\
DF\,Cap			& EC\,20338--1925 		& 15.7	& $35.5\pm0.3$ 	& $5.75\pm0.10$ 	& $-1.71\pm0.10$	& 135--168 	&  5 	& P	& 8 & 6 \\
DK\,Cap			& PHL\,44				& 13.2	& $26.6\pm1.1$	& $5.41\pm0.05$ 	& $-2.97\pm0.10$	& 3007--7763	& 9	& G	& 9, 10 & 11 \\
EO\,Cet     		& PB\,8783        	        & 12.3 	& $36.2\pm0.3$ 	& $5.70\pm0.05$  	& ... 				& 94--136 		& 11	& P	& 12 & 13 \\ 
HK\,Cnc			& PG\,0865+121		& 13.6	& $26.4\pm1.0$	& $5.73\pm0.15$	& $-3.00\pm4.34$	& 313--435	& 2	& P	& 14 & 15\\ 
LT\,Cnc			& PG\,0907+123		& 13.9	& $26.2\pm0.9$	& $5.30\pm0.10$ 	& $-1.74\pm0.21$	& 3477--7448	& 7	& G 	& 16 &  17, 18\\ 
AQ\,Col     		& EC\,05217--3914 		& 15.6 	& $31.3\pm0.3$ 	& $5.76\pm0.06$  	& ...				& 215--218 	& 8 	& P	& 19, 20 & 19 \\ 
\\
V2203\,Cyg  		& KPD\,2109+4401  		& 13.3 	& $31.8\pm0.6$ 	& $5.76\pm0.05$  	& $-2.23\pm0.10$	& 182--213 	& 8 	& P	& 21, 22 & 23\\
V2214\,Cyg  		& KPD\,1930+2752  		& 13.8 	& $33.8\pm0.3$ 	& $5.58\pm0.04$  	& $-1.52\pm0.02$	& 158--326 	& 68 	& P	& 24, 25  & 24, 26 \\
LM\,Dra      		& PG\,1618+563B   		& 13.5 	& $33.9\pm0.5$ 	& $5.80\pm0.09$ 	& $-1.60\pm0.10$	& 108--144 	&  2 	& P	& 27, 28 & 28 \\
LS\,Dra      		& HS\,1824+5745   		& 15.6 	& $31.5\pm0.7$ 	& $6.00\pm0.16$ 	& $-1.55\pm0.14$	&  139 		&  1 	& P	& 1, 5 & 1, 29 \\
V366\,Dra			& SBSS\,1716+581		& 16.7 	& $34.4\pm0.3$ 	& $5.75\pm0.10$ 	& $-1.80\pm0.10$	& 137--144 	&  2 	& P	& 30, 31 &  6 \\
V1078\,Her   		& PG\,1613+426    		& 14.4 	& $34.4\pm0.5$ 	& $5.97\pm0.12$  	& $-1.66\pm0.06$	&  144 		&  1 	& P	& 32 & 32\\
V1093\,Her		& PG\,1716+426		& 14.0	& $26.6\pm0.6$	& $5.42\pm0.12$ 	& $-2.89\pm0.20$	& 2939--5460	& 6	& G	& 33, 34 &  15, 35 \\
V1211\,Her		& SDSS\,J164214.21+425234.0 & 15.7 & $32.4\pm0.3$ & $5.80\pm0.10$	& $-2.00\pm0.10$	& 130--138 	&  2 	& P	& 7 &  6 \\
V361\,Hya      		& EC\,14026--2647 	        & 15.3 	& $35.0\pm0.4$ 	& $6.14\pm0.09$  	& ...				& 134--144 	&  3	& P	& 36 & 13 \\
V541\,Hya			& PG\,0958--116	        & 15.3 	& $34.8\pm0.2$ 	& $5.80\pm0.04$ 	& $-1.68\pm0.06$	& 136--169 	&  5 	& P	& 8 & 37 \\
\\
V551\,Hya			& EC\,11583--2708 		& 14.4 	&  ... 				& ...  				& ...				& 114--149 	&  4 	& P	& 8 & \\
DT\,Lyn     		& PG\,0911+456    		& 14.6 	& $31.9\pm0.1$ 	& $5.80\pm0.02$  	& $-2.55\pm0.06$	& 149--192 	&  7 	& P	& 38, 39 & 38, 39 \\
DV\,Lyn      		& HS\,0815+4243   		& 16.1 	& $33.7\pm0.3$ 	& $5.95\pm0.05$ 	& $-2.10\pm0.10$	& 126--131 	&  2 	& P	& 40 & 40 \\
DW\,Lyn   			& HS\,0702+6043   		& 14.3 	& $28.4\pm0.6$ 	& $5.35\pm0.10$ 	& $-2.70\pm0.10$ 	& 363--3538 	&  3 	& H	& 41, 42 & 41 \\
V2579\,Oph		& PG\,1627+017		& 12.9	& $23.7\pm0.2$	& $5.29\pm0.02$ 	& $-3.96\pm0.03$	& 4580--8901	& 23	& G	& 43, 44 & 35, 45 \\
V1405\,Ori  		& KUV\,04421+1416		& 15.1 	& $32.0\pm0.4$ 	& $5.72\pm0.10$  	& ...				& 156--242 	& 18	& P	& 19, 46 & 19 \\
V1636\,Ori   		& HS\,0444+0458   		& 15.4 	& $33.8\pm1.0$ 	& $5.60\pm0.15$ 	& $-1.85\pm0.20$	& 137--169 	&  3 	& P	& 1, 3 & 1 \\
V1835\,Ori		& RAT\,J0455+1305 		& 17.2 	& $29.2\pm1.9$ 	& $5.20\pm0.30$ 	& ...				& 184--4361 	& 12 & H	& 47, 48, 49 &48 \\ 
V384\,Peg    		& HS\,2149+0847   		& 16.5 	& $35.6\pm1.0$ 	& $5.90\pm0.20$ 	& $-1.80\pm0.20$	& 142--159 	&  6	& P	& 40, 50 & 40 \\
V387\,Peg    		& HS\,2151+0857   		& 16.5 	& $34.5\pm1.3$ 	& $6.10\pm0.25$ 	& $-1.37\pm0.20$ 	& 129--151 	&  5 	& P	&  1, 5 & 1 \\
\\
V391\,Peg    		& HS\,2201+2610   		& 13.6 	& $29.3\pm0.5$ 	& $5.40\pm0.10$  	& $-3.00\pm0.30$	& 344--3256 	&  7 	& H	& 51,  52, 53 & 40 \\ 
V585\,Peg			& Balloon\,090100001  	& 11.8 	& $29.4\pm0.5$ 	& $5.33\pm0.10$ 	& $-2.54\pm0.20$	& 118--10111	& 114& H	& 54, 55, 56, 57  & 54 \\
EK\,Psc      		& PG\,0014+067    		& 16.3 	& $34.5\pm2.7$ 	& $5.78\pm0.01$ 	& $-1.66\pm0.31$	& 77--173 		& 19 	& P	& 58, 59,60 & 58 \\
EP\,Psc      		& PG\,2303+019    		& 16.2 	& $35.2\pm1.5$ 	& $5.74\pm0.15$ 	& $-1.70\pm0.10$	& 128--145 	&  3 	& P	& 61 & 61 \\
EQ\,Psc			& PB\,5450			& 13.1	& ...				& ...				& ...				& 1246--10697	& 18	& G	& 62 \\
V338\,Ser   		& PG\,1605+072    		& 12.8 	& $32.3\pm0.3$ 	& $5.25\pm0.05$  	& $-2.53\pm0.10$	& 350--573 	& 50 	& P	& 63,  64 & 63, 65 \\
V499\,Ser			& SDSS\,J160043.60+074802.8 & 17.7& $69.3\pm1.4$	& $6.01\pm0.06$	& $-0.68\pm0.04$	& 60--119		& 10	& P	& 66 & 67, 68, 69 \\
UX\,Sex     		& EC\,10228--0905 		& 15.9 	& $34.4\pm0.6$ 	& $5.84\pm0.16$  	& ...				& 140--152 	&  3 	& P	& 70, 13 & 13 \\
UY\,Sex     		& PG\,1047+003    		& 13.5 	& $35.0\pm1.0$ 	& $5.90\pm0.10$  	& ...				& 104--175 	& 18 	& P	& 71, 72, 73 & 72 \\
V4640\,Sgr  		& EC\,20117--4014 		& 12.5 	& $34.9\pm0.3$ 	& $5.90\pm0.06$  	& ...				& 137--158 	&  3 	& P	& 13, 74 & 13 \\
\\
KL\,UMa     		& Feige\,48       		& 13.3 	& $29.5\pm0.2$ 	& $5.49\pm0.03$  	& $-2.94\pm0.04$	& 345--378 	&  4	& P	& 75, 76  & 75, 23, 77 \\
KY\,UMa     		& PG\,1219+534    		& 13.2 	& $33.7\pm0.3$ 	& $5.83\pm0.04$  	& $-1.49\pm0.06$	& 122--149 	&  7 	& P	& 38, 78 & 38, 23, 79 \\
NY\,Vir        		& PG\,1336--018   		& 13.5 	& $31.4\pm0.2$ 	& $5.62\pm0.04$  	& $-2.93\pm0.05$	&  97--205 	& 28 	& P	& 80, 81 & 80, 82 \\
QQ\,Vir      		& PG\,1325+101    		& 14.0 	& $35.3\pm0.1$ 	& $5.85\pm0.03$  	& $-1.70\pm0.02$	&  94--168 	& 14 	& P	& 61, 83  & 15, 72, 84 \\
V594\,Vir			& SDSS\,J144514.93+000249.0  & 17.4 & $35.9\pm0.3$& $5.75\pm0.10$ 	& $-1.60\pm0.10$	& 120--142 	&  3 	& P	& 7 &  6 \\
2MASS\,J04155016+0154209     & ...           	& 14.0 	& $34.0\pm0.5$ 	& $5.80\pm0.05$  	& $-1.60\pm0.05$	& 144--149 	&  3 	& P	& 85 & 85  \\
Balloon\,081400001	& BPS\,CS\,22890--0074	& 14.0	& $29.4\pm2.3$	& $5.60\pm0.30$ 	& ...				& 3142--10286	& 8	& G	& 86 & 86 \\
CS\,1246        		& CS\,124636.2--631549 	& 14.6 	& $28.5\pm0.7$ 	& $5.46\pm0.11$ 	& $-2.00\pm0.30$	&  372 		&  1 	& P	& 87 & 87  \\
EC\,00404--4429	& GSC\,07538--00411	& 13.7	& ...				& ...				& ...				& 7194--9774	& 4	& G	& 88 &  \\
EC\,01541--1409      & GD\,1053           		& 12.2 	& $37.1\pm0.3$ 	& $5.71\pm0.10$ 	& $-1.71\pm0.10$	&  64--164 	& 34 	& P	& 89, 90 & 6 \\
\\
EC\,15094--1725	& ...					& 16.8	& ...				& ...				& ...				& 129--145	& 3	& P	& 88 & \\
EC\,22221--3152      & GSC 07497-00351		& 13.4 	& $35.1\pm0.3$ 	& $5.85\pm0.10$ 	& $-1.47\pm0.10$	&  84--176 	& 10 	& P	& 89 &  6 \\ 
EC\,22590--4819	& LB\,1516			&13.0	& $25.9\pm0.8$	& $5.45\pm0.08$ 	& $-2.75\pm0.09$	& 3116--20870	& 8	& G	& 91 & 92, 93, 94 \\
EPIC\,203948264	& ...					& 16.7	& ...				& ...				& ...				& 1800--10080	& 17	& G	& 95 & \\
FBS\,0117+396		& ...					& 15.3	& $28.7\pm0.4$	& $5.48\pm0.06$	& $-3.05\pm0.10$	& 337--8696	& 10	& H 	& 96, 97 & 98, 97 \\ 
GALEX\,J080656.7+152718 & ...			& 14.2	& $28.5\pm0.4$	& $5.36\pm0.05$ 	& $-2.96\pm0.01$	& 354--2417	& 4	& H	& 99 & 100, 99 \\
HE\,0218--3437	& ...					&13.4	& $25.3\pm0.3$	& $5.40\pm0.10$	& ...				& 4105--8204	& 2	& G	& 91 & 93 \\
HE\,0230--4323  	& ...              			& 13.8 	& $31.6\pm0.5$ 	& $5.60\pm0.07$ 	& $-2.58\pm0.06$	& 282--310 	&  5 	& P	& 101, 102 & 103 \\ 
HE\,1450--0957  	& EC\,14507--0957		& 15.3 	& $34.6\pm0.5$ 	& $5.79\pm0.07$ 	& $-1.29\pm0.06$	& 118--139 	&  3 	& P	& 6 & 103  \\
HE\,2151--1001  	& ...           			& 15.6 	& $35.0\pm0.5$ 	& $5.70\pm0.07$ 	& $-1.60\pm0.06$ 	& 126--128 	&  2 	& P	& 6 & 103\\
\\
HE\,2316--0909	& PHL\,457			& 13.0	& $28.4\pm0.2$	& $5.52\pm0.02$ 	& $-2.46\pm0.07$	& 3766--16615	& 3	& G	& 88 & 92, 104, 11\\
HS\,2125+1105   	& ...           			& 16.3 	& $32.5\pm0.5$ 	& $5.76\pm0.07$ 	& $-1.86\pm0.06$	& 136--146 	&  2 	& P	& 6 & 103 \\
JL\,82			& EC\,21313--7301		& 12.4	& $25.4\pm0.3$	& $5.11\pm0.07$ 	& ...				& 3305--8438	& 13	& G	& 105 & 93, 106 \\ 
JL\,166	        		& GSC\,08022--01020       & 15.0 	& $34.6\pm0.9$ 	& $5.80\pm0.02$ 	& $-1.08\pm0.21$	&  97--167 	& 10 	& P	& 107 & 92,  107 \\ 
JL\,194			& HIP\,2499			& 12.4	& $25.2\pm1.2$	& $5.20\pm0.20$ 	& $-1.82\pm1.54$	& 4375--7855	& 5	& G	& 16 & 108 \\
{\bf{J0902--0720}}	& {\bf TYC\,4890--19--1}			& {\bf 12.4}	& {$\mathbf{34.2\pm0.5}$}	& {$\mathbf{5.87\pm0.10}$} 	& {$\mathbf{-1.38\pm0.10}$}	& {\bf 136--140}		& {\bf 2}	& {\bf P}	& {\bf109} & {\bf109} \\
{\bf{J1938+5609}}	& {\bf 2MASS\,J19383247+5609446}& {\bf 13.3	}	& {$\mathbf{29.5\pm0.5}$}	& {$\mathbf{5.34\pm0.10}$} 	& {$\mathbf{-2.86\pm0.10}$}	& {\bf 373	}		& {\bf 1}	& {\bf P} 	& {\bf109} & {\bf109} \\
{\bf{J2344--3427}}	& {\bf HE\,2341--3443}			& {\bf 11.0	}	& {$\mathbf{28.0\pm0.3}$}	& {$\mathbf{5.40\pm0.20}$} 	& {$\mathbf{-2.98\pm0.14}$}	& {\bf 387--9954}	& {\bf 4}	& {\bf H}	& {\bf109} & {\bf108, 110, 11} \\
KIC\,1718290		& ...					& 15.5	& $21.8\pm0.1$	& $4.67\pm0.03$ 	& $-0.40\pm0.04$	& 4607--40306	& 56	& (G)& 111 & 111 \\
KIC\,2437937		& NGC\,6791\,B5		& 18.0	& $24.7\pm0.6$	& $5.50\pm0.07$ 	& $-2.67\pm0.21$	& 3201--8607	& 4	& (G)& 112 & 113, 112 \\
KIC\,2438324		& NGC\,6791\,B4		& 18.3	& $26.2\pm0.4$	& $5.54\pm0.06$ 	& $-2.86\pm0.02$	& 2385--7641	& 19	& G	& 114, 115 &  113, 112 \\
\hline
\end{tabular}
\end{table*}

\begin{table*}
\sevensize
%\scriptsize    
\contcaption{Catalogue of the sdV stars. }
\label{tab:sdBV-cat_cont}
  \begin{tabular}{llccccccccc}
    \hline

	\multicolumn{1}{l}{Name} & \multicolumn{1}{l}{Alternative} & $m_V$ & $T_{\rm eff}$ & $\log g$ 			& $\log(N_{\rm He}/N_{\rm H})$ &Period range & N$_P$ & Pulsator 	& \multicolumn{2}{c}{Reference} \\
		                                  & \multicolumn{1}{l}{name}        &             & (kK)               &   (cm\,s$^{-2}$)     	&            					 &(s)                 &             &    Type 	& Pulsation 	& Parameters\\
		  \hline
KIC\,2569576		& NGC\,6791\,B3		& 17.8	& $24.5\pm0.4$	& $5.28\pm0.04$ 	& $-2.76\pm0.12$	& 3284--7919	& 11	& G	& 112 & 113, 112 \\ 
KIC\,2697388		& ...					& 15.4	& $23.9\pm0.3$	& $5.32\pm0.03$ 	& $-2.90\pm0.10$	& 263--14493	& 59 & H	& 116, 117, 118 & 119 \\
KIC\,2991276		& ...					& 17.4	& $33.9\pm0.2$	& $5.82\pm0.04$ 	& $-3.10\pm0.10$	& 118--215	& 8	& P	& 119, 120 & 119 \\
KIC\,2991403		& ...					& 17.4	& $27.3\pm0.2$	& $5.43\pm0.03$ 	& $-2.60\pm0.10$	& 2710--12760	& 38	& G	& 119, 115 & 119 \\
KIC\,3527751		& ...					& 14.9	& $27.9\pm0.1$	& $5.35\pm0.03$ 	& $-2.98\pm0.04$	& 270--13889	& 251& G (H)	& 116,  121 & 119, 121 \\
KIC\,5807616		& KPD\,1943+4058		& 14.9	& $27.3\pm0.2$	& $5.52\pm0.02$ 	& $-2.85\pm0.08$	& 2346--9126	& 26	& G	& 116, 122 & 119, 122 \\
KIC\,7664467		& ...					& 16.5	& $26.8\pm0.5$	& $5.17\pm0.08$	& $-2.80\pm0.20$	& 4050--9076	&  6	& (G)	& 116 & 119 \\
KIC\,7668647 		& FBS\,1903+432		& 15.5	& $27.7\pm0.3$	& $5.48\pm0.02$ 	& $-2.64\pm0.03$	&  211--28508	& 132 & (H)	& 123, 124 & 125, 124 \\
KIC\,8302197		& ...					& 16.4	& $27.2\pm0.2$	& $5.39\pm0.02$ 	& $-2.61\pm0.06$	& 166--13889	& 31	&  (H)	& 123, 126, 127 & 125, 127 \\
\\
KIC\,9472174		& TYC\,3556--3568--1	& 12.7	& $29.6\pm0.1$	& $5.41\pm0.01$ 	& $-2.40\pm0.10$	& 220--19884	& 55	& P (H)	& 128 & 119  \\
KIC\,10001893		& ...					&15.8	& $26.7\pm0.3$	& $5.30\pm0.04$ 	& $-2.90\pm0.10$	& 341--12899	& 27	& G (H)	& 126 & 125 \\
KIC\,10139564		& ...					& 16.1	& $32.0\pm0.1$	& $5.71\pm0.02$ 	& $-2.21\pm0.03$	& 123--5263	& 60 & P (H)	& 129, 130 &119, 130  \\
KIC\,10553698		& ...					& 15.1	& $27.5\pm0.2$	& $5.42\pm0.02$ 	& $-2.81\pm0.02$	& 246--21349	& 43 & G (H)	&  126, 131 &125, 131 \\
KIC\,10670103		& ...					& 16.5	& $21.1\pm0.3$	& $5.12\pm0.03$ 	& $-2.59\pm0.04$	& 1485--43478	& 28	& G		& 116, 132 & 119, 132 \\
KIC\,11179657		& ...					& 17.1	& $26.0\pm0.8$	& $5.14\pm0.13$ 	& $-2.10\pm0.20$	& 2844--5362	& 11	& G		& 119, 129 & 119 \\
KIC\,11558725		& ...					& 14.9	& $27.7\pm0.1$	& $5.40\pm0.01$ 	& $-3.12\pm0.02$	& 198--26415	& 166& (H) 	& 126, 133 & 119, 133 \\
KPD\,0629-0016	& ...					& 14.9	& $27.4\pm0.3$	& $5.50\pm0.02$ 	& $-2.76\pm0.04$	& 2601--10516	& 17	& G		& 134, 135 & 134, 136 \\
LS\,IV--14\,116		& ...					& 12.9	& $34.7\pm0.1$	& $5.87\pm0.02$ 	& $-0.62\pm0.01$	& 1954--2870	& 2	& G		& 137 & 138, 139, 140, 141 \\
PB\,7032			& BPS\,CS\,22965--0031	& 13.2	& $27.7\pm0.2$	& $5.50\pm0.03$ 	& ...				& 1807--17109	& 12	& G		& 86 &  86 \\
\\
PG\,0048+091		& EPIC\,220614972		& 14.3 	& $34.2\pm0.3$ 	& $5.69\pm0.10$ 	& $-3.00\pm0.10$	&  90--192 	& 28 	& P		& 4, 142 &  6 \\
PG\,0101+039		& Feige\,11			& 12.1	& $28.3\pm0.1$	& $5.53\pm0.02$ 	& $-2.76\pm0.03$	& 2650--7235	& 3	& G		& 143 & 15, 143 \\
PG\,1033+201    	& ...           			& 15.4 	& ... 				& ...  				& ...				& 146--171 	&  2 	& P		&   6 &  \\ 
PG\,1338+481		& ...					& 13.7	& $28.2\pm0.3$	& $5.38\pm0.04$ 	& ...				& 2125--9510	& 14	& G		& 144 & 144 \\
PG\,1142-037		& EPIC\,201206621		& 15.8	& $28.0\pm0.1$ 	& $5.31\pm0.01$ 	& $-2.87\pm0.03$	& 3362--9199	& 14	& G		& 145 & 145 \\
PG\,1657+416    	& ...           			& 16.2 	& $32.2\pm0.5$ 	& $5.73\pm0.10$ 	& $-2.03\pm0.15$	& 125--143 	&  5 	& P		& 146 & 146 \\
SDSS\,J233406.10+462249.3 	& ...			& 17.7	& $34.6\pm0.5$	& $5.71\pm0.09$	& $-1.30\pm0.10$	& 128--134	& 2	& P		& 147 & 98 \\
TYC\,1077--218--1	& GSC\,01077--00218	& 12.2	& $32.1\pm1.0$	& $5.15\pm0.20$ 	& $-2.80\pm0.20$	& 376--565	& 16	& P		& 148 & 148 \\
TYC\,3389--882--1	& 2MASS\,J06395217+5157013 & 12.0& $30.8\pm0.3$	& $5.70\pm0.08$ 	& $-2.59\pm0.29$	& 260--1744	& 2	& H		& 149 & 100 \\
V2008-1753		& ...					& 16.7	& $32.8\pm0.3$ 	& $5.83\pm0.04$ 	& $-2.27\pm0.13$	& 152--182	& 4	& P		& 150 & 150 \\
\\
$\omega$\,Cen\,V1	& ...					& 18.5	& $48.5\pm0.6$	& $5.80\pm0.05$ 	& $-1.82\pm0.06$	& 85--115	     	& 2	& P		& 151 & 152 \\
$\omega$\,Cen\,V2	& ...					& 18.4	& $49.9\pm0.9$	& $5.54\pm0.05$ 	& $-1.84\pm0.07$	& 102--108	& 2	& P		& 151 & 152 \\
$\omega$\,Cen\,V3	& ...					& 18.5	& $49.3\pm0.9$	& $6.01\pm0.07$ 	& $-1.72\pm0.10$	& 103--110	& 2	& P		& 151, 153, &152  \\
$\omega$\,Cen\,V4	& ...					&  18.4	& $52.0\pm0.7$	& $5.83\pm0.73$ 	& $-1.24\pm0.05$	& 114--124	& 2	& P		& 153 & 152\\
$\omega$\,Cen\,V5	& ...					& 18.7	& $53.4\pm1.2$	& $6.10\pm0.09$ 	& $-1.66\pm0.13$	& 99--108		& 2	& P		& 152 &152 \\
NGC\,2808\,VAR1	& EHB3				& 18.2$^a$& ...				& ...				& ...				& 108--116	& 2	& P		& 154 &\\
NGC\,2808\,VAR2	& ...					& 18.3$^a$& ...				& ...				& ...				& 150		& 1	& P		& 154 &\\
NGC\,2808\,VAR3	& ...					& 18.1$^a$& ...				& ...				& ...				& 121		& 1	& P		& 154 &\\
NGC\,2808\,VAR4	& ...					& 18.1$^a$& ...				& ...				& ...				& 85--104		& 1	& P		& 154 &\\
NGC\,2808\,VAR5	& BHk1				& 18.4$^a$& ...				& ...				& ...				& 147		& 1	& P		& 154 &\\
NGC\,2808\,VAR6	& BHk6				& 18.3$^a$& ...				& ...				& ...				& 113		& 1	& P		& 154 &\\
\hline
\multicolumn{11}{l}{$^a$NUV magnitude from \citet{brown01}}\\
\multicolumn{11}{l}{References: 1) \citet{ostensen01a}; 2) \citet{jeffery04}; 3) \citet{reed07}; 4) \citet{koen04}; 5) \citet{reed06}; 6) \citet{ostensen10b};}\\

\multicolumn{11}{l}{7) \citet{solheim06}; 8) \citet{kilkenny06}; 9) \citet{kilkenny06c}; 10) \citet{kilkenny07}; 11) \citet{geier13b}; 12) \citet{odonoghue98b};}\\
\multicolumn{11}{l}{13) \citet{odonoghue97}; 14) \citet{piccioni00}; 15) \citet{saffer94}; 16) \citet{koen10b}; 17) \citet{kupfer15}; 18) \citet{drilling13};}\\
\multicolumn{11}{l}{19) \citet{koen99a}; 20) \citet{reed07b}; 21) \citet{billeres98}; 22) \citet{zhou06}; 23) \citet{heber00}; 24) \citet{billeres00};} \\
\multicolumn{11}{l}{25) \citet{reed11}; 26) \citet{geier06}; 27) \citet{schuh00}; 28) \citet{silvotti00}; 29) \citet{edelmann03}; 30) \citet{solheim04}; 31) \citet{aerts06};}\\
\multicolumn{11}{l}{32) \citet{bonanno03}; 33) \citet{green03}; 34) \citet{reed04b}; 35) \citet{morales03}; 36) \citet{kilkenny97}; 37) \citet{randall09};}\\
\multicolumn{11}{l}{38) \citet{koen99b}; 39) \citet{randall07}; 40) \citet{ostensen01b}; 41) \citet{dreizler02}; 42) \citet{schuh06}; 43) \citet{randall04};}\\

\multicolumn{11}{l}{44) \citet{randall06}; 45) \citet{for06}; 46) \citet{reed10b}; 47) \citet{ramsay05}; 48) \citet{ramsay06}; 49) \citet{baran11e};}\\
\multicolumn{11}{l}{50) \citet{kilkenny06b}; 51) \citet{silvotti02}; 52) \citet{lutz09}; 53) \citet{silvotti10}; 54) \citet{oreiro04}; 55) \citet{baran05}; 56) \citet{oreiro05};}\\ 
\multicolumn{11}{l}{57) \citet{baran09}; 58) \citet{brassard01}; 59) \citet{jeffery05}; 60) \citet{vuckovic06}; 61) \citet{silvotti02b}; 62) \citet{jeffery14};}\\
\multicolumn{11}{l}{63) \citet{koen98a}; 64) \citet{kilkenny99}; 65) \citet{heber99}; 66) \citet{woudt06}; 67) \citet{fontaine08}; 68) \citet{rodriguez10};}\\
\multicolumn{11}{l}{69) \citet{latour11}; 70) \citet{stobie97a}; 71) \citet{billeres97}; 72) \citet{odonoghue98}; 73) \citet{kilkenny02}; 74) \citet{randall06c};}\\
\multicolumn{11}{l}{75) \cite{koen98b}; 76) \citet{reed04}; 77) \citet{charpinet05b}; 78) \citet{reed09}; 79) \citet{charpinet05c}; 80) \citet{kilkenny98};}\\

\multicolumn{11}{l}{81) \citet{kilkenny03}; 82) \citet{vuvckovic07}; 83) \citet{silvotti06}; 84) \citet{charpinet06}; 85) \citet{oreiro09}; 86) \citet{koen11}; 87) \citet{barlow10};}\\
\multicolumn{11}{l}{88) \citet{kilkenny16}; 89) \citet{kilkenny09}; 90) \citet{reed12}; 91) \citet{koen10c}; 92) \citet{heber86}; 93) \citet{edelmann03a}; 94) \citet{chayer06};}\\
\multicolumn{11}{l}{95) \citet{ketzer16}; 96) \citet{geier12b}; 97) \citet{ostensen13}; 98) \citet{geier11}; 99) \citet{baran11b}; 100) \citet{vennes11};}\\
\multicolumn{11}{l}{101) \citet{koen07}; 102) \citet{kilkenny10}; 103) \citet{lisker05}; 104) \citet{blanchette08}; 105) \cite{koen09}; 106) \citet{geier10};}\\
\multicolumn{11}{l}{107) \citet{barlow09}; 108) \citet{heber84}; 109) this work; 110) \citet{nemeth12}; 111) \citet{ostensen12}; 112) \citet{reed12b};}\\

\multicolumn{11}{l}{113) \citet{liebert94}; 114) \citet{pablo11}; 115) \citet{baran12b}; 116) \citet{reed10}; 117) \citet{charpinet11};}\\
\multicolumn{11}{l}{118) \citet{baran12c}; 119) \citet{ostensen10c}; 120) \citet{ostensen14}; 121) \citet{foster15}; 122) \citet{vangrootel10}; 123) \citet{reed11b};}\\
\multicolumn{11}{l}{124) \citet{telting14}; 125) \citet{ostensen11b}; 126) \citet{baran11}; 127) \citet{baran15}; 128) \citet{ostensen10d}; 129) \citet{kawaler10};}\\

\multicolumn{11}{l}{130) \citet{baran13}; 131) \citet{ostensen14b}; 132) \citet{reed14}; 133) \citet{telting12}; 134) \citet{koen07b}; 135) \citet{charpinet10};}\\
\multicolumn{11}{l}{136) \citet{vangrootel10b}; 137) \citet{ahmad05}; 138) \citet{viton91}; 139) \citet{ahmad03}; 140) \citet{green11}; 141) \citet{randall15};}\\
\multicolumn{11}{l}{142) \citet{reed07a}; 143) \citet{randall05}; 144) \citet{randall06b}; 145) \citet{reed16}; 146) \citet{oreiro07}; 147) \citet{pakvstiene14};}\\ 
\multicolumn{11}{l}{148) \citet{ostensen11}; 149) \citet{vuckovic12}; 150)\citet{schaffenroth15}; 151) \citet{randall10}; 152) \citet{randall16};}\\
\multicolumn{11}{l}{153) \citet{randall11}; 154) \citet{brown13}}
\end{tabular}
\end{table*}

As is seen with stars on the main-sequence, there is a final subset of sdV stars which show both low- and high-frequency pulsations. These hybrid pulsators (also called DW\,Lyn stars after their prototype), first seen by \citet{schuh06}, provide the ideal laboratory to test the internal structure of the subdwarf stars. The presence of both p and g\,modes in a single star allows for detailed asteroseismological modelling of the interior as the different modes probe varying layers in the star. For example, to explain the presence of both modes in a single star, \citet{jeffery06} were able to extend the instability strip of the sdBV stars, so that the p- and g-mode regions overlapped, by including nickel opacities in their calculations. 

The original differences in pulsation frequency which were used to identify p-mode, g-mode and hybrid pulsators is not robust enough when presented with precise space-based observations. Therefore, in Table\,\ref{tab:sdBV-cat}, we class the spaced-based observations as if they were observed from the ground, with a detection limit of $1$\,mmag. If there are other pulsations below this limit, a classification of `(H)' is given to indicate that the star is a hybrid pulsator at the detection limit of the {\it Kepler} satellite. This notation is also used in the cases where all pulsations are below the ground-based detection limit, i.e. in the cases where a classification of `(G)' is given.

\begin{figure}
  \centering
  \includegraphics[width=\linewidth]{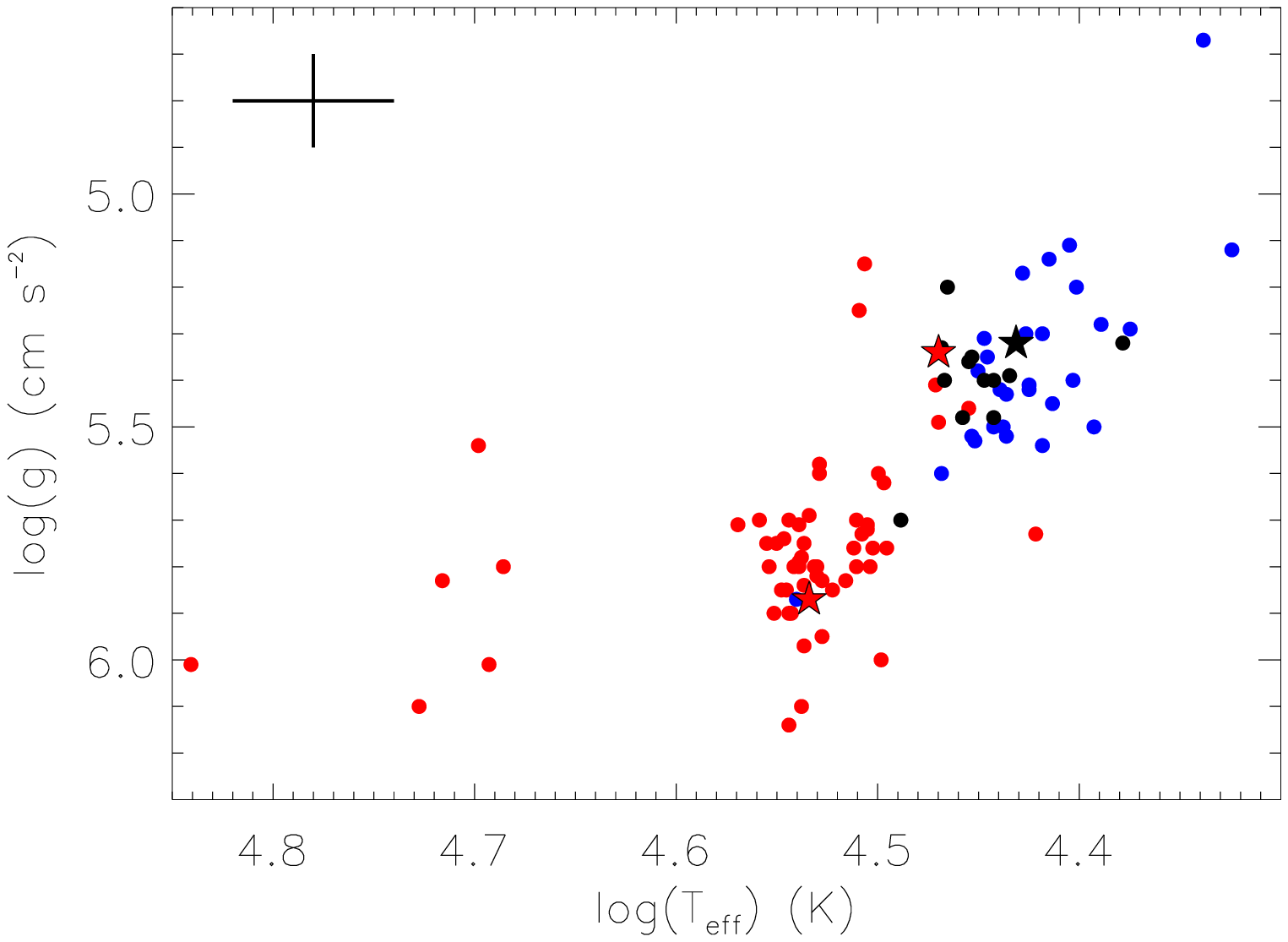}
  \includegraphics[width=\linewidth]{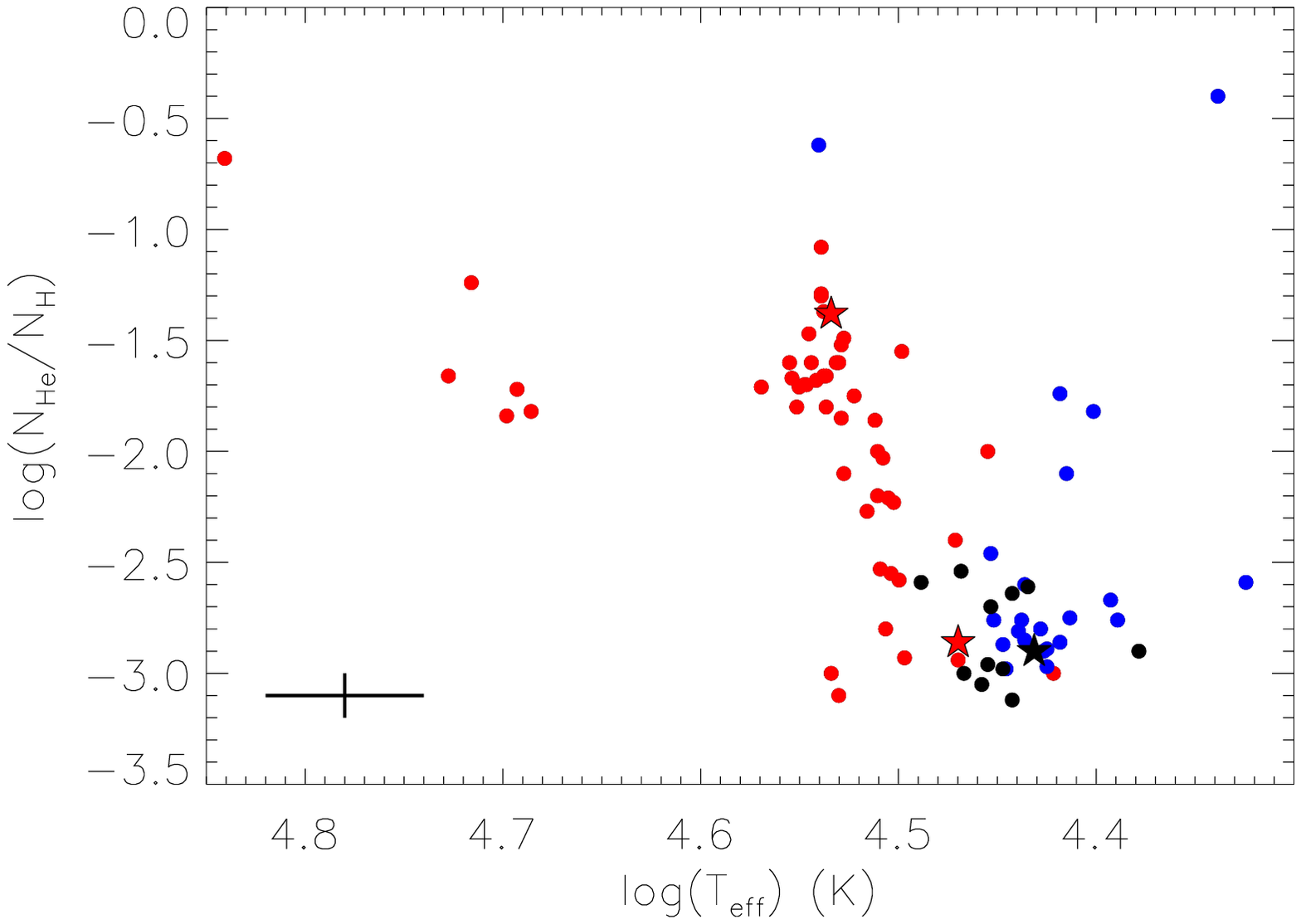}
  \includegraphics[width=\linewidth]{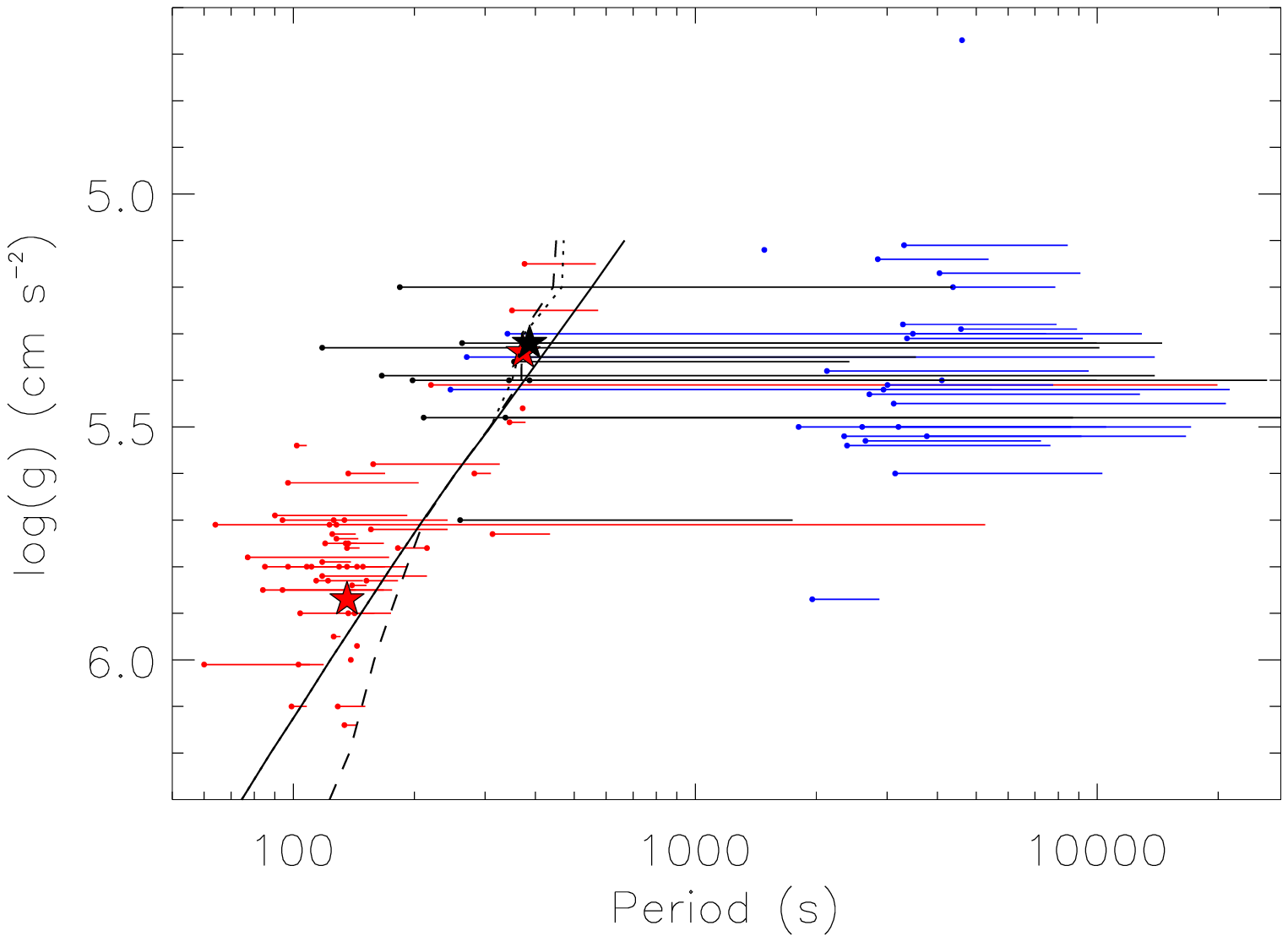}
  \caption{Top: the known sdV stars in the $T_{\rm eff}-\log g$ plane for which data is available in Table\,\ref{tab:sdBV-cat}. The red dots indicate the short period p-mode pulsators, the blue dots are the long period g-mode pulsators and the black dots are the hybrid pulsators. The classifications are assigned using the ground-based criterion if available,  otherwise the space-based classification. The targets we report here are shown by the outlined stars. Middle: the $T_{\rm eff}-\log(N_{\rm He}/N_{\rm H})$ diagram for the stars in Table\,\ref{tab:sdBV-cat}. In general, the hotter stars are the p-mode pulsators and are more He-rich. Bottom: the $P-\log g$ diagram for the same stars. The horizontal lines emanating from dots represent the range of the pulsation periods for a given star. The black lines represent the loci of models for the $\ell=0, n=0$ mode (solid line), the $\ell=2, n=0$ mode (dashed line) and the $\ell=1,n=1$ mode (dotted line) for the p-mode pulsators. See \citet{koen99b} for a discussion of the models. The black crosses in the top two panels indicate the typical errors. See online version for colour plots.}
  \label{fig:teff_logg}
\end{figure}

The top panel of Fig.\,\ref{fig:teff_logg} shows all the known sdV stars, which have values of $T_{\rm eff}$ and $\log g$ in Table\,\ref{tab:sdBV-cat}, in the $T_{\rm eff}-\log g$ plane. Two groups form in the diagram, the blue points which represent the g-mode pulsators, and the red points which represent the p-mode pulsators. The g-mode pulsators have systematically lower temperatures and surface gravities than their p-mode counterparts. The hybrid stars, as one would expect, straddle the two groups, and are shown in black. One star of particular note is the g-mode pulsator LS\,IV--14\,116, the blue dot amongst the red. This star shows an extremely peculiar abundance pattern and is thought to belong to the halo population \citep{randall15}. LS\,IV--14\,116 is also an outlier in the middle and bottom panels of Fig.\,\ref{fig:teff_logg} where the abnormal helium abundance is obvious and it is the only high surface gravity g-mode pulsator. Models of this star, using the opacity mechanism, cannot explain the pulsations given the derived stellar parameters. Despite considerable effort, this star remains a mystery \citep[e.g.][]{miller11,naslim11,green11}.

The middle panel of Fig.\,\ref{fig:teff_logg} shows the effective temperature against helium abundance for the stars in Table\,\ref{tab:sdBV-cat}. The hotter stars tend to have a higher He abundance than their cooler counterparts, as expected \citep[e.g.][]{edelmann03}, and as such are predominantly the p-mode pulsators. There are two stars that do not follow the general trend in the plot: the g-mode pulsator LS\,IV--14\,116 which has both high helium abundance and a high temperature, and the cool g-mode pulsator KIC\,1718290. The former, as described above, has an extremely peculiar helium abundance. The latter, KIC\,1718290, was shown to be on the Blue Horizontal branch, rather than the EHB, by \citet{ostensen12} which may explain its obviously different position in the effective temperature against helium abundance plot when compared to the other g-mode pulsators.

Finally, the bottom panel of Fig.\,\ref{fig:teff_logg} shows the range of pulsation period of a star against its surface gravity. Models have shown \citep[e.g.][]{fontaine98,koen99a} that there is an expected relationship between the pulsations in the sdBV stars and their surface gravity. The black lines plotted represent the fundamental modes of $\ell=0,n=0$  (solid line) and $\ell=2,n=0$ (dashed line), and the $\ell=1,n=1$ (dotted line) mode for p-mode pulsators. Most stars lie on the short period side of the solid line, with some at higher surface gravities better constrained by the dashed line. It must be noted, however, that surface gravity determinations in the sdBV stars can be greatly influenced by binarity, metallicity and the pulsations, forcing models to fit higher surface gravities if the effects are not accounted for.

Ground-based observations of the g-mode and hybrid pulsators can be challenging due to the intrinsic low amplitudes and low observational duty cycles. As such, observations by the {\it Kepler} Space Telescope have pushed the study of sdBV stars into a new age with the observations of many low-amplitude frequency-rich stars \citep[e.g.][]{ostensen10d,ostensen10c,baran11,baran12}. The precision and time-base of the observations of these stars will not be surpassed for many years. 

However, with the upcoming Transiting Exoplanet Survey Satellite \citep[TESS;][]{ricker15} mission, many of the bright stars detected with ground-based observations will be revisited and be subject to $\umu$mag precision observations for days to months at a time. TESS fields will be observed for 27-d at a minimum cadence of 30\,min, with 750 and 60 targets observed at 120\,s and 20\,s, respectively, per field. Where fields overlap, multiple 27-d periods will combine to provide longer time-base observations, and in the best cases up to 1-yr at the ecliptic poles. As such, the identification of further, bright, targets is key to expanding the sample size available for TESS observations which will have a limiting magnitude of $I_C\lesssim10-13$ \citep{ricker15}. 

\subsection{The SuperWASP project}

The Super Wide Angle Search for Planets (WASP) project is a two-site wide-field survey for transiting exoplanets \citep{pollacco06}. The instruments are located at the Observatorio del Roque de los Muchachos on La Palma and at the Sutherland Station of the South African Astronomical Observatory (SAAO). The instruments consist of eight 200\,mm, f/1.8 Canon telephoto lenses backed by Andor CCDs of $2048\times2048$ pixels, allowing a pixel size of about 14\,arcsec. Observations are made through broadband filters of $4000-7000$\,\AA. An instrument reconfiguration was conducted in 2012 on SuperWASP-S. This involved changing the 200\,mm lenses for 85\,mm to enable the targeting of brighter stars for planetary transits \citep{turner15}. 

The data pass through a custom reduction pipeline correcting for primary and secondary extinction, the colour-response of the instrument, the zero point and atmospheric extinction. The pipeline is optimised for G stars. The data are also corrected for instrumental systematics using the {\sc{sysrem}} algorithm of \citet{tamuz05}. The observing strategy of SuperWASP provides two consecutive 30\,s exposures at a given pointing, before moving to the next available field; fields are typically revisited every 10\,min. Such an observing strategy allows for a nominal Nyquist frequency of 1440\,\cd, but in reality due to the pseudo-random sampling there is no strict Nyquist frequency. Frequency analysis is only limited by the length of the exposure.

The SuperWASP project has been shown to have a versatile archive in which to search for a whole host of variable stars \citep[e.g.][]{maxted08,thomas10,norton11,smalley11,smalley14b,smalley17,mcquillin12,holdsworth14a,holdsworth15}. In this paper, we provide an in depth analysis of three sdBV stars which have been identified to vary by \citet{holdsworth14a} and \citet{holdsworth15}. These stars were not previously known to be pulsating sdB stars, and as such, provide further examples of the sdBV group of variable stars. These targets are prime candidates for follow-up observations by the TESS space mission due to there relative brightness among the sdV stars.

Throughout this paper, the reader must bear in mind that the amplitudes presented are those detected in the broadband filter of the SuperWASP instrument. As such, they will be greatly reduced when compared to other sdV stars which are typically observed through narrower band filters in the blue part of the spectrum. The pulsation amplitudes in these stars are greater in the blue part of the spectrum as a result of their high temperature and the variations in the effective temperature over the pulsation cycle.

\section{The targets}

\subsection{J1938+5609}

\begin{figure}
  \centering
  \includegraphics[width=\linewidth]{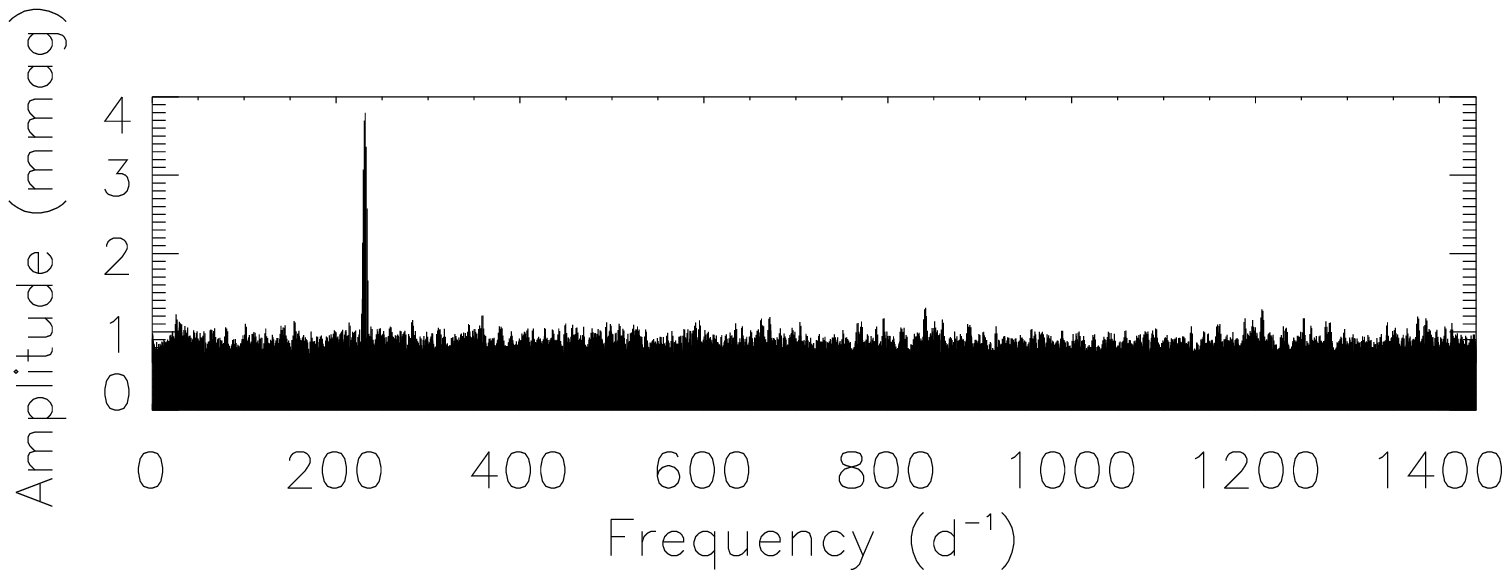}
\includegraphics[width=\linewidth]{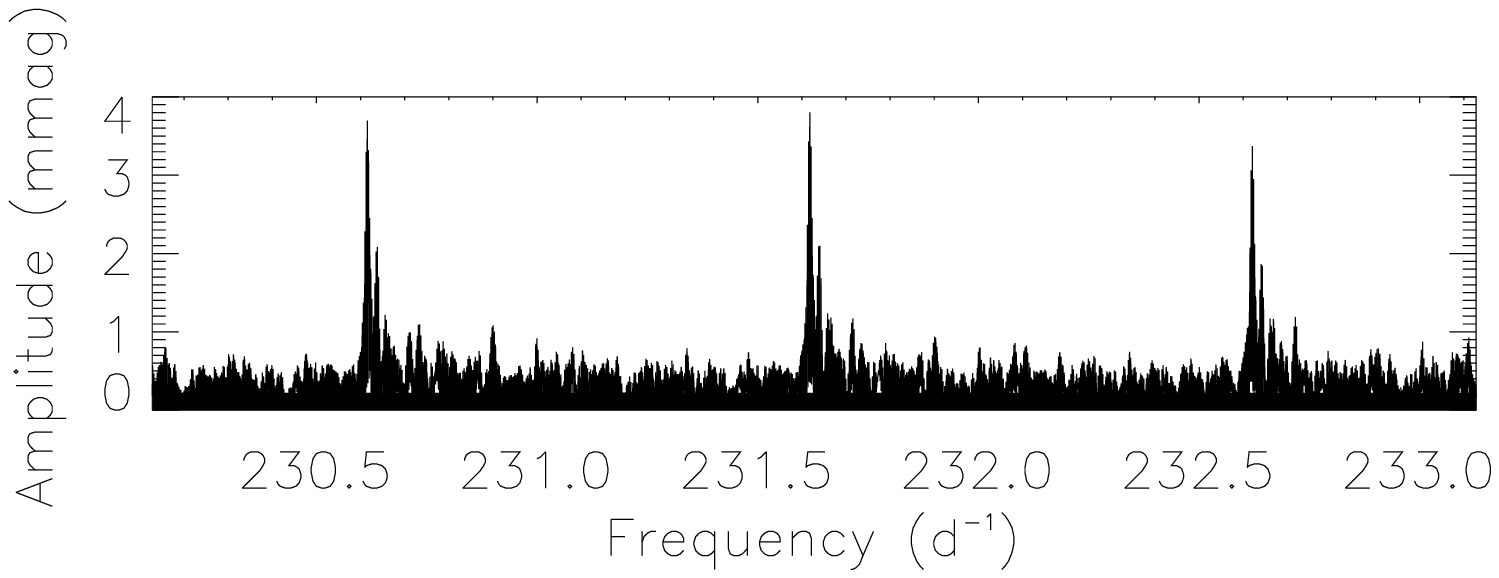}  
  \caption{Top: amplitude spectrum of the discrete Fourier transform of all data, to the nominal Nyquist frequency, for J1938+5609 showing periodic variability at 231\,\cd. Bottom: zoomed view of the variability and the alias structure. The second peak can be seen on the high-frequency shoulder of the main signal.}
  \label{fig:J1938-ft}
\end{figure}

J1938+5609 ($\alpha$:\,19:38:32.48, $\delta$:\,+56:09:44.6) is a newly identified sdB star which pulsates at a frequency of 231.62\,\cd\, ($P=373$\,s) with an average amplitude of 4\,mmag (Fig.\,\ref{fig:J1938-ft}). The star was observed by SuperWASP over a period of 4 yr $(2007-2010)$, and was observed simultaneously by two different cameras in 2007 and 2008. Table\,\ref{tab:J1938-wasp} details the SuperWASP observations. Where there are multiple observations per season, a letter is added to the season column to aid differentiation. Peaks which had the greatest amplitude in the combined data set were taken to be the true variability of the star. The peaks which surround the peak at greatest amplitude are a result of the window pattern, which is dominated by the daily aliases which plague ground-based, single-site, time-resolved observations.

\begin{table*}
    \caption{Details of the SuperWASP observations for J1938+5609, and the results of a non-linear least-squares fitting to each of the seasons. Time (BJD) is given as BJD-240\,0000.0. The zero point for the phases is taken to be the centre point of each of the data sets.}
\label{tab:J1938-wasp}
  \begin{tabular}{lcrccccrr}
    \hline
    Season & BJD & \multicolumn{1}{c}{Length} & Number of & \multicolumn{1}{c}{ID} &Frequency & Amplitude & \multicolumn{1}{c}{Phase} & S/N\\
           & start & \multicolumn{1}{c}{(d)}  & points                     &     &  (\cd)      & (mmag)    & \multicolumn{1}{c}{(rad)} & \\
    \hline
      2007a & 4230.5894 & 66.0356 & 2658 & $\nu_1$ & $231.6179\pm0.0009$ & $6.61\pm0.66$ & $0.106 \pm0.113$ & 6.37 \\
      2007b & 4249.5562 & 46.1206 & 2383 & $\nu_1$ & $231.6209\pm0.0021$ & $4.68\pm0.82$ & $-1.817\pm0.178$ & 4.11 \\
      2008a & 4631.7129 & 58.6768 & 1545 & $\nu_1$ & $231.6195\pm0.0019$ & $7.98\pm1.69$ & $0.322 \pm0.217$ & 3.30 \\
      2008b & 4605.6377 & 84.9810 & 5732 & $\nu_1$ & $231.6179\pm0.0008$ & $5.14\pm0.54$ & $-1.249\pm0.109$ & 6.10 \\
    2009   & 4971.6289 & 127.8379 & 8489 & $\nu_1$ & $231.6189\pm0.0005$ & $3.89\pm0.39$ & $1.627 \pm0.098$ & 6.70 \\
  	        &		   &		     &		 & $\nu_2$ & $231.6405\pm0.0007$ & $2.53\pm0.38$ & $-2.991\pm0.154$ & 4.57 \\
  2010   & 5336.6289 & 122.8848 & 10748 & $\nu_1$ & $231.6174\pm0.0006$ & $3.13\pm0.39$ & $2.164 \pm0.123$ & 5.88 \\
        	       &		   &		     &		 & $\nu_2$ & $231.6397\pm0.0008$ & $2.49\pm0.38$ & $-1.879\pm0.155$ &5.12 \\
      All & 4230.5894 & 1228.9243 & 31555 & $\nu_1$ & $231.61808\pm0.00002$ & $3.77\pm0.23$ & $-0.608\pm0.062 $ & 11.08 \\
        	       &		   &		     &		 & $\nu_2$ & $231.64058\pm0.00004$ & $1.98\pm0.23$ & $-0.914\pm0.118$ & 6.34 \\
	           \hline
  \end{tabular}
\end{table*}

To confirm this star as an sdB star, an optical spectrum of J1938+5609 was obtained with Intermediate dispersion Spectrograph and Imaging System (ISIS) instrument mounted on the 4.2-m William Herschel Telescope (WHT) on 2015 Mar 12. We used the R600B grating with a 1.5\,arcsec slit, attaining a resolution of R$\sim$2\,000. The exposure time was 1000\,s, leading to a peak S/N$\sim$180. The spectrum has been reduced in the standard way, including flat-field correction, de-biasing and wavelength calibrations applied. Tools from the {\sc starlink} project\footnote{http://starlink.eao.hawaii.edu/starlink} were used to perform these tasks. The spectrum was intensity rectified using the {\sc uclsyn} spectral synthesis package \citep{smalley01}.

\begin{figure}
  \centering
  \includegraphics[width=\linewidth]{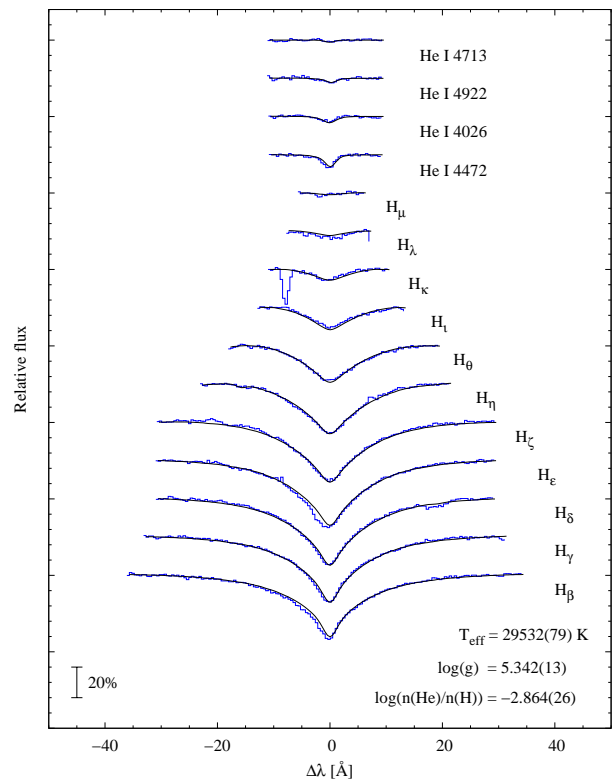}
  \caption[Spectrum of J1938+5609]{Fitting results of the WHT/ISIS spectrum of J1938+5609 confirming it to be an sdB star. Values in brackets of the fit parameters are the errors in the last digits, see the text for more realistic errors. The artefact in the blue wing of H$_\kappa$ is from a bad CCD column.}
  \label{fig:J1938-spec}
\end{figure}

The spectrum is quite featureless, with only Balmer lines and He\,{\sc{i}} lines present. The hydrogen and helium lines of the extracted spectrum were fitted to a grid of synthetic spectra calculated from fully line blanketed LTE model atmospheres assuming solar metallicity \citep{heber00}. The result is shown in Fig.\,\ref{fig:J1938-spec}. The artefact in the blue wing of H$_\kappa$ is from a bad CCD column. Note that formal fitting errors stated in the figure do not account for systematic effects inherent in the models, so we generously increase the errors when stating $T_{\rm eff}=29\,500\pm500$\,K, $\log g=5.34\pm0.10$\,cm\,s$^{-2}$, and $\log(N_{\rm He}/N_{\rm H})=-2.86\pm0.10$. 

To analyse the light curve, we treat each season of data separately. To remove the remaining low-frequency `red' noise from the light curve after the data have been processed by the SuperWASP pipeline, we prewhiten the data to 10\,\cd\, to an amplitude limit which is representative of the noise level at high-frequency. This frequency limit is sufficiently removed from the pulsation as to not affect the subsequent analysis. This is an iterative process where we identify, fit and remove peaks above the high-frequency noise level. We then apply linear and non-linear least-squares fitting to the light curve to extract the frequency of variability. The results of the non-linear least-squares fitting are shown in Table\,\ref{tab:J1938-wasp}. 

\begin{figure}
  \centering
  \includegraphics[width=\linewidth,angle=180]{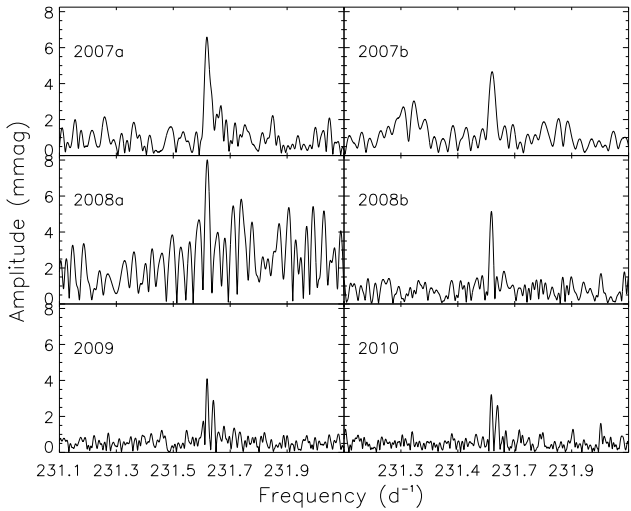}
  \caption[Periodogram of the separate seasons for J1938+5609]{Amplitude spectrum of the discrete Fourier transform of the separate seasons for J1938+5609 showing the amplitude variation of the peak, and the splitting of the mode into two well resolved frequencies.}
  \label{fig:J1938-montage}
\end{figure}

As can be seen from the results, the amplitude of the pulsation changes over the 4-yr period. This is shown graphically in Fig.\,\ref{fig:J1938-montage}. The 2008a season of data is shorter in duration and low on the number of points, making the amplitude determination much more difficult, as is demonstrated by the large error in Table\,\ref{tab:J1938-wasp}. The other seasons show, on average, a decrease in the pulsation amplitude of the principal peak, and an eventual emergence of a second peak at $\nu_2=231.64$\,\cd.

The presence of the second peak in the later seasons, combined with the decrease in amplitude, suggests there may be beating between the two peaks or with further, unresolved, peaks. However, the separation of the two peaks, $\delta\nu=0.022\pm0.001$\,\cd, is equal to the lowest frequency resolution of the SuperWASP data (namely the 2007b data). As such, the two peaks should be resolved in {\emph{all}} the periodograms. The low quality of the WASP data do not allow us to confirm the presence of variable amplitude over the course of the observations presented here -- although the results are suggestive of variability, the noise and resolution are not sufficient for its presence to be confirmed.

Very few sdBV stars show just one or two pulsation modes (see Table\,\ref{tab:sdBV-cat}). Other than the stars identified by \citet{brown13} for which no accurate temperatures are published, J1938+5609 is the second coolest single mode variable. The temperature and surface gravity of J1938+5609 place it among the hybrid pulsators (see Fig.\,\ref{fig:teff_logg}). In terms of its stellar parameters this target is similar to Balloon\,090100001 \citep{baran08,telting08,baran09}, which shows amplitude variations in both photometric and spectroscopic observations. The variations seen in the amplitude spectrum of Balloon\,090100001 are likely to be caused by energy transfer between the p and g\,modes. We therefore postulate this might be the case for J1938+5609, and that there maybe further modes present in this star, but below our detection limit.

Independently of the SuperWASP survey, J1938+5609 was targeted as part of the survey for bright pulsators in the GALEX sample, which also yielded J20133+0928 \citep{ostensen11}, J08069+1527 \citep{baran11b}, and J06398+5156 \citep{vuckovic12}. All  observations were obtained with the Merope\,{\sc ii} frame-transfer imager on the 1.2-m Mercator telescope \citep{ostensen10a}. The data were processed by standard overscan bias-level subtraction and flat-fielding, and the light curves extracted by aperture photometry, using the Real Time Photometry ({\sc{rpt}}) program \citep{ostensen01b}. Observations were made through two different filters: the R and B filters of the Geneva system denoted RG and BG, respectively. The targets was first observed on the night of 2011 Jun 7 during two short runs. Those discovery runs were immediately followed up with a long run on 2011 Jun 10. Another fairly long run was obtained on the night of 2011 Aug 13 during the same campaign as for FBS\,0117+396 \citep{ostensen13}, while waiting for that target to become high enough to observe. Finally, it was observed on the nights of 2011 Sept 3 and 4, where the observations on the first of those nights suffered from intermittent clouds. 

\begin{figure}
	\centering
	\includegraphics[width=\linewidth]{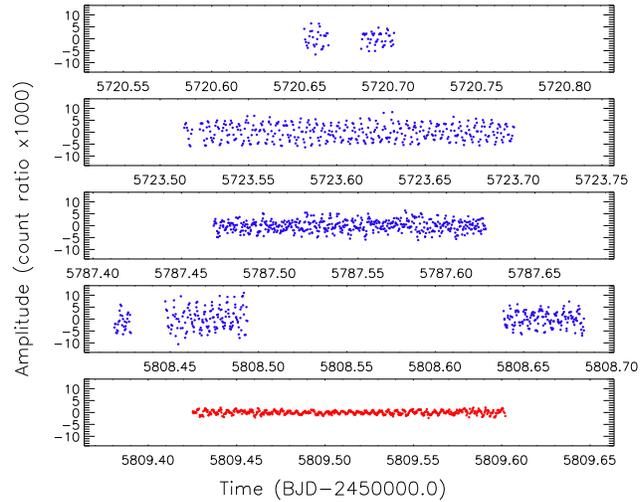}
	\caption{Mercator light curve of J1938+5609. Note the variable amplitude seen in the data. The observations in the top four panels, blue data points, were obtained through a BG filter, with the red data points in the bottom panel obtained through an RG filter. See the online version for colour plots.}
	\label{fig:MRC_LC}
\end{figure}

The extracted light curves are shown in Fig.\,\ref{fig:MRC_LC}. The star shows amplitude variability over the period of the observations. However, the data are not of sufficient length to resolve more than one mode, as seen in SuperWASP. The manner in which the data are extracted with the {\sc rpt} programme provides a magnitude in counts relative to the comparison stars. As such, the extraction of amplitude from the data for comparison to the SuperWASP white light data is not possible. The data do, however, demonstrate the amplitude suppression as a function of filter response, and thus the need for blue-band observations of these stars to maximise frequency detection.

\subsection{J0902--0720}

\begin{figure}
  \centering
  \includegraphics[width=\linewidth]{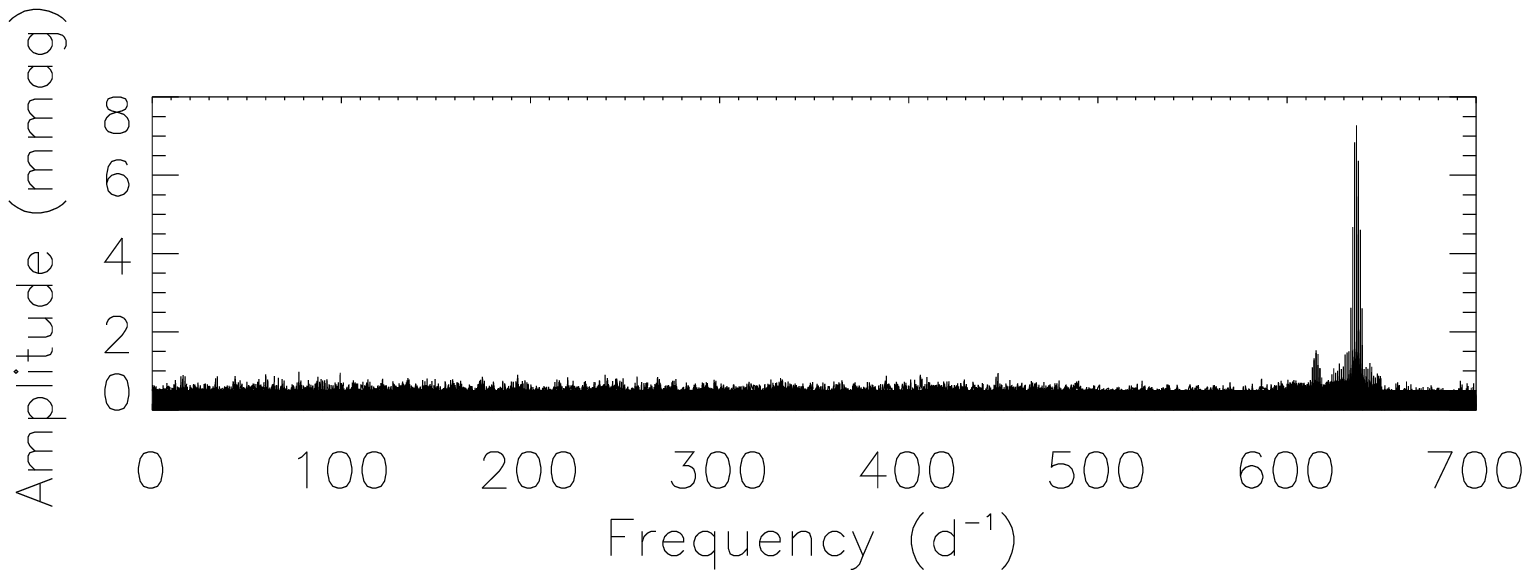}
  \includegraphics[width=\linewidth]{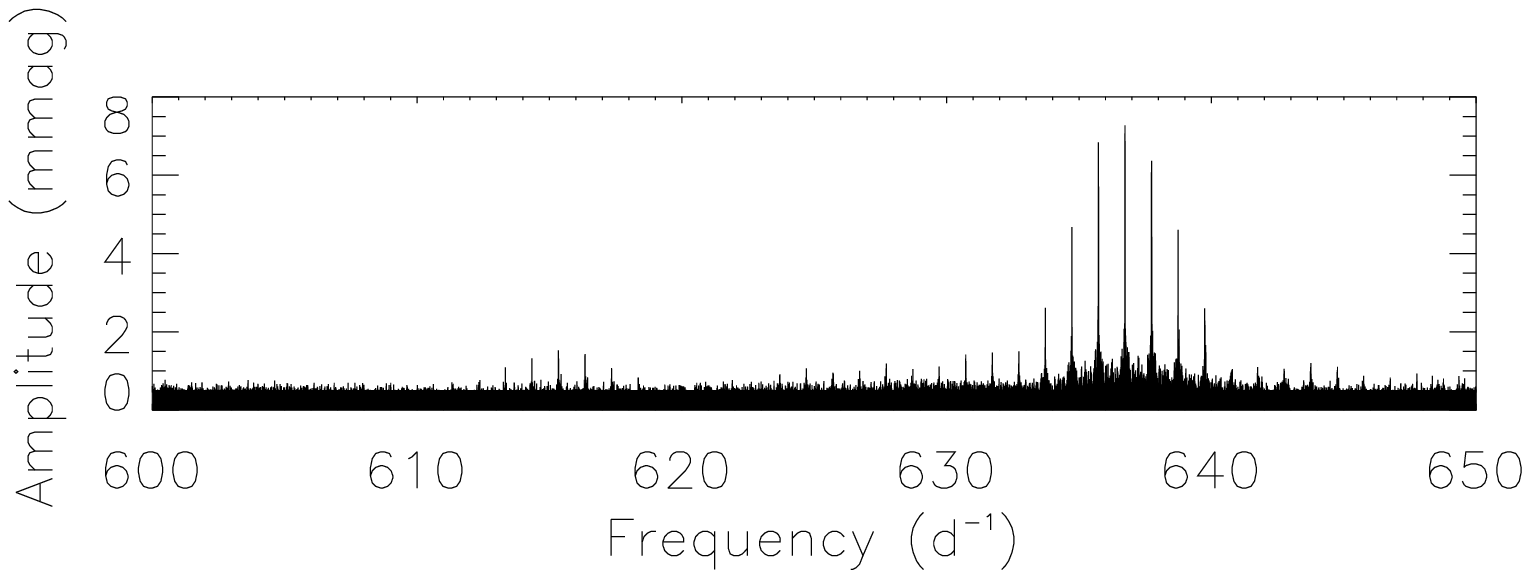}
  \caption{Top: full amplitude spectrum of the discrete Fourier transform of the SuperWASP data of J0902--0720 showing two periodic variations. Bottom: zoomed view of the variations seen in the top panel.}
  \label{fig:J0902-ft}
  \end{figure}

J0902--0720 ($\alpha$\,: 09:02:04.52, $\delta$:\,$-07$:20:47.58; TYC 4890-19-1) is a newly identified sdB star, detected through the light variations in SuperWASP data. The data show variability at 636.74\,\cd\, ($P=136$\,s) with an average amplitude of 7.27\,mmag, and a second peak at 615.34\,\cd\, ($P=140$\,s) with an average amplitude of 1.53\,mmag (Fig.\,\ref{fig:J0902-ft}). 

\begin{figure}
  \centering
  \includegraphics[width=\linewidth]{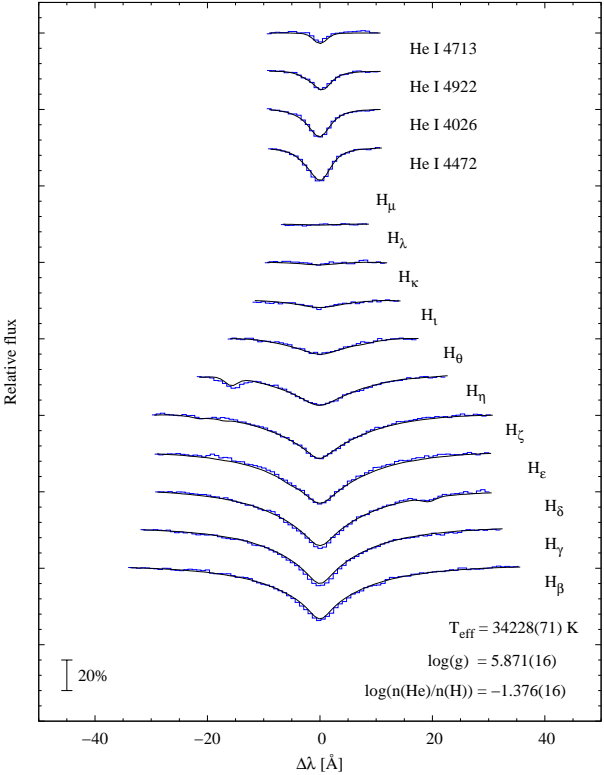}
  \caption[Spectrum of J0902]{The results of spectral fitting of the NOT/ALFOSC spectrum of J0902--0720. Values in brackets of the fit parameters are the errors in the last digits. See the text for more realistic errors.}
  \label{fig:J0902-spec}
\end{figure}

 A low-resolution spectrum (R$\sim$1360) of J0902--0720 was obtained with the Andalucia Faint Object Spectrograph and Camera, ALFOSC, instrument mounted on the 2.56-m Nordic Optical Telescope, on 2013 Feb 04. We used grism \#16 and a 1.0 arcsec slit. The exposure time was 300\,s, leading to a peak S/N$\sim$220 in the extracted spectrum. Standard reduction steps within {\sc iraf} include bias subtraction, removal of pixel-to-pixel sensitivity variations, optimal spectral extraction, and wavelength calibration based on arc-lamp spectra. There are many He lines in the spectrum, with very few other features (excluding the H lines). The results of the spectral fitting are shown in Fig.\,\ref{fig:J0902-spec}, where the errors shown are the formal fitting errors. When taking into account systematic effects, we adopt $T_{\rm eff}=34\,200\pm500$\,K, $\log g = 5.87\pm0.10$\,cm\,s$^{-2}$ and $\log(N_{\rm He}/N_{\rm H}) = -1.38\pm0.1$, confirming this star as a pulsating subdwarf B star. These parameters place the star amongst the p-mode pulsators.

To analyse the light curve of J0902--0720, we prewhitened each season of data to 10\,\cd\, to an amplitude representative of the high-frequency noise, following the procedure described earlier. We then conducted linear and non-linear least-squares fitting of the pulsations. The results of the non-linear least-squares fitting are shown in Table\,\ref{tab:J0902-wasp}. As before, seasons with multiple data are differentiated by a letter, and peaks are identified from the aliases in the amplitude spectrum of the combined data.

\begin{table*}
    \caption{Details of the SuperWASP observations for J0902--0720, and the results of a non-linear least-squares fitting to each of the seasons. Time (BJD) is given as BJD-240\,0000.0. The zero point for the phases is taken to be the centre point of each of the data sets.}
\label{tab:J0902-wasp}
  \begin{tabular}{lcrcccrrr}
    \hline
    Season & BJD & \multicolumn{1}{c}{Length} & Number of & \multicolumn{1}{c}{ID} &\multicolumn{1}{c}{Frequency} & \multicolumn{1}{c}{Amplitude} & \multicolumn{1}{c}{Phase} & S/N\\
           & start & \multicolumn{1}{c}{(d)}  & points                     &     &  \multicolumn{1}{c}{(\cd)}      & \multicolumn{1}{c}{(mmag)}    & \multicolumn{1}{c}{(rad)} \\
    \hline
    2009a & 4846.3565 & 97.0020 & 4652 	  & $\nu_1$ & $636.7340\pm0.0002$ & $12.14\pm0.50$ & $-1.184\pm0.041$ & 9.15 \\
     							& & & & $\nu_2$ & $615.3372\pm0.0010$ & $2.67\pm0.50$   & $0.373 \pm0.191$ & 4.16 \\
     							& & & & $\nu_3$ & $615.4361\pm0.0012$ & $2.19\pm0.50$   & $-1.107\pm0.233$ & 3.84 \\
    
    2009b & 5167.6401 & 118.8760 & 2908  & $\nu_1$ & $636.7356\pm0.0002$ & $9.59\pm0.47$   & $2.092 \pm0.051$ & 7.75 \\
     							& & & & $\nu_2$ & $615.3393\pm0.0011$ & $1.91\pm0.48$   & $-2.868\pm0.256$ & 3.22 \\
     
    2009c & 5168.6729 & 117.8433 & 3388  & $\nu_1$ & $636.7355\pm0.0002$ & $10.69\pm0.42$  & $0.710\pm0.042$ & 8.23 \\
     							& & & & $\nu_2$ & $615.3406\pm0.0009$ & $2.42 \pm0.42$   & $1.315\pm0.186$ & 3.95 \\

    2010a & 5212.4082 & 95.9702 & 4119 	  & $\nu_1$ & $636.7358\pm0.0003$ & $11.77\pm0.52$ & $1.845 \pm0.044$ & 8.95 \\
                        					 & & & & $\nu_2$ & $615.3403\pm0.0015$ & $2.11 \pm0.54$ & $-2.229\pm0.246$ & 3.93 \\
      							 & & & & $\nu_3$ & $615.4376\pm0.0012$ & $2.62 \pm0.54$ & $-1.404\pm0.203$ & 3.66 \\
    
    2010b & 5532.7612 & 116.7583 & 3759   & $\nu_1$ & $636.7371\pm0.0002$ & $9.60\pm0.35$  & $0.058 \pm0.038$ & 8.34 \\
     							  & & & & $\nu_2$ & $615.3411\pm0.0007$ & $2.44\pm0.35$ & $-2.013\pm0.148$ & 5.06 \\
     
    2010c & 5543.5894 & 105.9302   & 3735 & $\nu_1$ & $636.7375\pm0.0002$ & $10.53\pm0.37$ & $2.208\pm0.035$ & 8.72 \\
     							  & & & & $\nu_2$ & $615.3407\pm0.0008$ & $2.28 \pm0.37$ & $1.212\pm0.163$ & 4.54 \\
     
    2011a & 5624.4238 & 51.9502    & 1908 & $\nu_1$ & $636.7384\pm0.0007$ & $12.49\pm0.79$ & $0.841 \pm0.064$ & 7.24 \\
    
    2011b & 5913.4404 & 125.9028  & 3387 & $\nu_1$ & $636.7385\pm0.0002$ & $12.49\pm0.58$ & $1.456 \pm0.046$ & 10.16\\
     						 	 & & & & $\nu_2$ & $615.3409\pm0.0011$ & $2.58 \pm0.58$ & $-0.354\pm0.227$ & 3.85 \\
							 
 All & 4846.3565 & 1192.9868& 27586 & $\nu_1$ & $636.737210\pm0.00001$ & $7.27\pm0.18$ & $0.435\pm0.025$ & 13.41\\
 						& & & & $\nu_2$ &  $615.339722\pm0.00006 $ & $1.59\pm0.18$ & $-3.023\pm0.116$ & 6.26\\
 						& & & & $\nu_3$ &  $615.435681\pm0.00009$ & $0.94\pm0.18$ & $-0.149\pm0.197$ & 3.71\\
    
    \hline
  \end{tabular}
\end{table*}

We detect the principal peak in all available seasons of SuperWASP data. The second peak, $\nu_2$ at 615.34\,\cd, is detected in all bar one season. That season, 2011a, has the shortest length and a noise level of about 2.4\,mmag at the frequency of the undetected peak. A third peak, $\nu_3$ at 615.44\,\cd, is detected in two of the seasons. This has a similar amplitude to $\nu_2$ and as such should be detected in most of the data sets. Further data are required to confirm $\nu_3$ as a real signal.

Assuming that the peaks seen in the periodogram of J0902--0720 at 136\,s are typical of those found in other sdBV stars, we conclude that the light variations are due to p-mode pulsations driven by the iron opacity bump \citep{ostensen10a}.

It must be noted here that the amplitudes of the pulsations in J0902--0720 will be reduced as a result of the relatively long exposure time of the SuperWASP observations, when compared to the pulsation period. Taking this into account, the amplitude is reduced by 8\,per\,cent in the SuperWASP passband. This is calculated using the relation
\begin{equation}
\frac{A}{A_0}={\rm sinc}\frac{\pi T_{\rm exp}}{P_{\rm puls}},
\end{equation}
where $A$ is the measured amplitude, $A_0$ is the intrinsic amplitude (in the observed passband), $T_{\rm exp}$ is the exposure time which in the case of SuperWASP is $30$\,s, and $P_{\rm puls}$ is the pulsation period.

The spectroscopic parameters derived for J0902--0720 place it among the p-mode pulsators (Fig.\,\ref{fig:teff_logg}). The temperature and surface gravity suggest no g\,modes would likely be present in the star. Further to this, the He abundance we derive for J0902--0720 is high. When compared to the other sdBV stars (see middle panel of Fig.\,\ref{fig:teff_logg}), there are few p-mode pulsators which show a higher He abundance. The position which J0902--0720 occupies in that diagram is similar to 2M0415+0154, HE\,1450-0957 and J23341+4622 all of which show just two or three pulsation modes. However, we cannot rule out the presence of further modes in J0902--0720 from our broadband photometry. This bright ($V=12.4$) target is ideal for dedicated follow-up observations. 

\subsection{J2344--3427}

J2344--3427 ($\alpha$:\, 23:44:22.01, $\delta$:\, $-34$:27:00.40; HE\,2341-3443) was included in the survey of \citet{ostensen10b}, but no significant pulsations were detected in that short run. With only $\sim800$ data points, the detection limit was just $\sim3$\,mmag, so their null detection is not in conflict with the SuperWASP detection presented here.

The SuperWASP observations of J2344--3427 cover five seasons, with a total of 43\,572 data points. To analyse the high-frequency variability in this star, we prewhiten each individual light curve to a frequency of 10\,\cd\, and an amplitude equalling the approximate noise level of the high-frequency range, as performed for the previous targets. When analysing the peaks in the low-frequency regime, we do not prewhitten the data -- as the peaks are close to the red noise peaks in frequency space, altering the light curve at low-frequency to remove noise may affect the intrinsic variability signal.

\begin{figure}
  \centering
  \includegraphics[width=\linewidth]{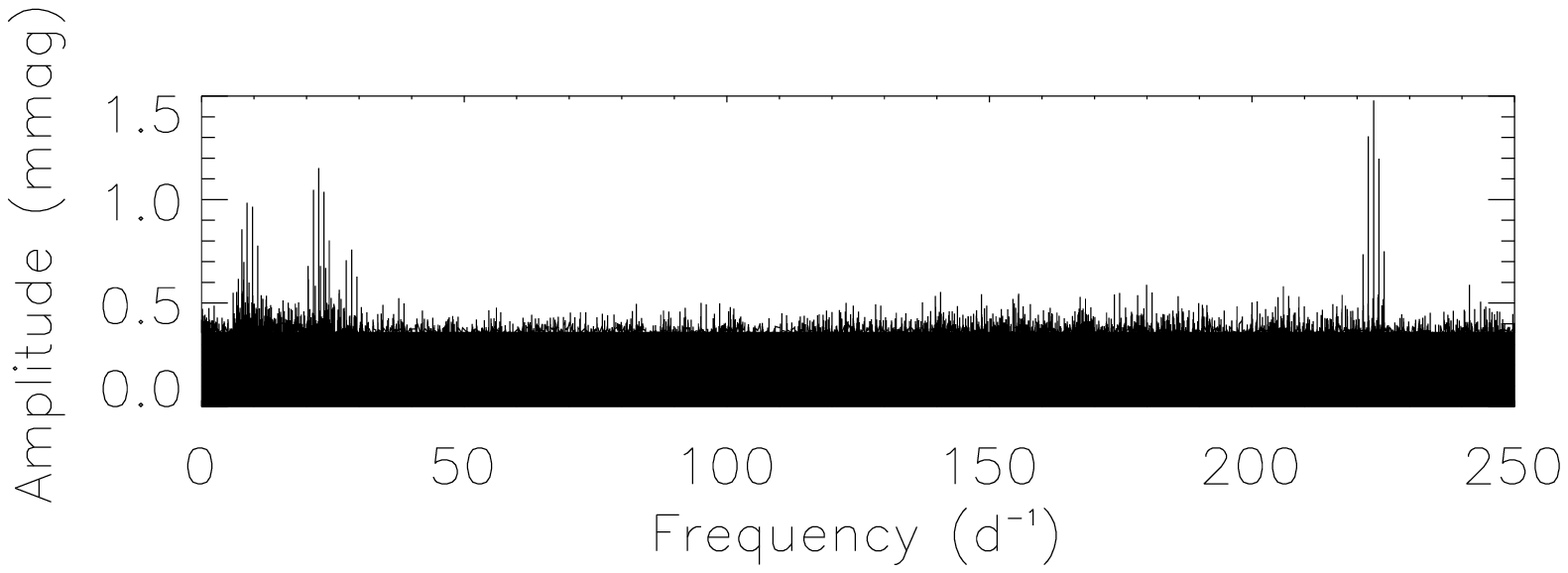}
  \includegraphics[width=\linewidth]{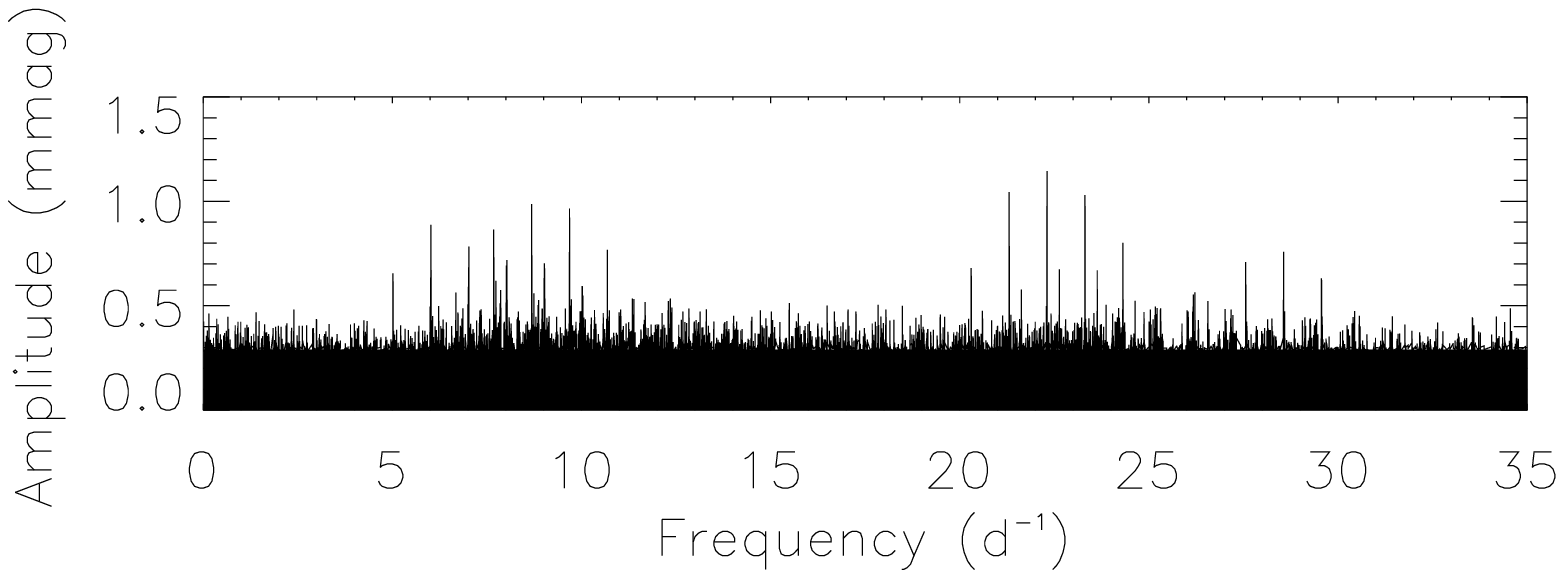}
  \caption{Top: full amplitude spectrum of the discrete Fourier transform of the SuperWASP light curve of J2344--3427 using the first three seasons of data. Bottom: the low-frequency range shown in more detail due to the presence of multiple frequencies.}
  \label{fig:J2344-ft}
\end{figure}

The high-frequency variation, at 223.16\,\cd, ($P=387$\,s) is detected in all seasons of the data with an average amplitude of 1.5\,mmag (Fig.\,\ref{fig:J2344-ft} top). As well as the high-frequency variability, there are further frequencies at 8.68, 22.31 and $28.56$\,\cd\, ($P=9954$, 3873 \& 3025\,s) with amplitudes of 0.99, 1.17 and 0.76\,mmag, respectively (Fig.\,\ref{fig:J2344-ft} bottom). These low-frequency variations are only detected (with significance) in the first three seasons of data. The lack of detections in the 2012 and 2013 seasons is most likely a result of the change to 85\,mm lenses, which results in a lower photometric precision per observation, and hence a higher noise level in the periodogram. The 2006, 2007 and 2011 seasons achieve a noise level of $\sim0.7$\,mmag in the low-frequency range, whereas the 2012 and 2013 data reach just 1.4\,mmag, greater than the detected pulsation amplitudes. A full frequency analysis, as well as a log of the SuperWASP observations, is shown in Table\,\ref{tab:J2344-wasp}.

\begin{table*}
    \caption{Details of the SuperWASP observations for J2344--3427, and the results of a non-linear least-squares fitting to each of the seasons. Time (BJD) is given as BJD-240\,0000.0. The zero point for the phases is taken to be the centre point of each of the data sets.}
\label{tab:J2344-wasp}
  \begin{tabular}{lclccrcrr}
    \hline
    Season & BJD & \multicolumn{1}{c}{Length} & Number of & \multicolumn{1}{c}{ID} &\multicolumn{1}{c}{Frequency} & Amplitude & \multicolumn{1}{c}{Phase} & S/N\\
           & start & \multicolumn{1}{c}{(d)}  & points                     &     &  \multicolumn{1}{c}{(\cd)}      & (mmag)    & \multicolumn{1}{c}{(rad)} \\
    \hline
      2006 & 3870.6143 & 183.8618 & 4574 & $\nu_1$ & $223.1601\pm0.0003$ & $1.46\pm0.16$ & $1.240 \pm0.096$ & 6.78 \\
	      &			   &  		      &		 & $\nu_2$ & $8.6803    \pm0.0004$ & $1.42\pm0.16$ & $2.972 \pm0.113$ & 5.14 \\
	      &			   &  		      &		 & $\nu_3$ & $22.3065  \pm0.0004$ & $1.37\pm0.16$ & $-2.851\pm0.116$ & 5.17 \\
	      &			   &  		      &		 & $\nu_4$ & $28.5596  \pm0.0006$ & $0.92\pm0.16$ & $2.432 \pm0.176$ & 4.48 \\
      
      2007 & 4268.5259 & 164.9025 & 4558 & $\nu_1$ & $223.1599\pm0.0004$ & $1.40\pm0.23$ & $-1.695\pm0.122$ & 4.98 \\
	      &			   &  		      &		 & $\nu_2$ & $8.6818    \pm0.0003$ & $2.32\pm0.23$ & $-1.892\pm0.101$ & 4.33 \\
	      &			   &  		      &		 & $\nu_3$ & $22.3071  \pm0.0011$ & $0.75\pm0.23$ & $2.978 \pm0.311$ & 2.54 \\ 
   	      &			   &  		      &		 & $\nu_4$ &  $28.5617 \pm0.0009$ & $0.86\pm0.23$ & $2.451 \pm0.269$ & 2.83 \\
	      
      2011 & 5731.5244 & 153.9233 & 3860 & $\nu_1$ & $223.1602\pm0.0004$ & $1.78\pm0.21$ & $-1.056\pm0.108$ & 5.96 \\
	      &			   &  		      &		 & $\nu_2$ & $8.3442    \pm0.0009$ & $0.91\pm0.21$ & $2.196 \pm0.230$ & 3.08\\
	      &			   &  		      &		 & $\nu_3$ & $22.3074  \pm0.0005$ & $1.50\pm0.21$ & $-0.279\pm0.138$ & 5.15 \\
	      &			   &  		      &		 & $\nu_4$ & $28.5623  \pm0.0010$ & $0.77\pm0.21$ & $-0.502\pm0.269$ & 2.92 \\
  
      2012 & 6111.5190 &  148.8671& 17174& $\nu_1$ & $223.1622\pm0.0011$ & $1.44\pm0.39$ & $2.091\pm0.291$  & 2.87 \\

      2013 & 6449.5532 & 187.7925 & 13406& $\nu_1$ & $223.1601\pm0.0012$ & $1.31\pm0.42$ & $2.549\pm0.332$ & 2.41 \\
      
      All & 3870.6143 & 2766.7314 & 43572 & $\nu_1$ & $223.16024\pm0.00003$ & $1.47\pm0.21$ & $0.561\pm0.169$  & 5.14 \\
	      &			   &  		      &		 & $\nu_3$ & $22.30749  \pm0.00003$ & $1.38\pm0.21$ & $1.459\pm0.197$ & 5.15 \\

    \hline
  \end{tabular}
\end{table*}

These low-frequency pulsations may originate from a main-sequence companion, as most sdB stars are found in binary systems, or they may be g\,modes in the same star, as is observed with the DW Lyn sdBV stars \citep{schuh06}. However, we are able to rule-out some binary scenarios: the lack of a detected orbital period, or its harmonic, allows us to exclude a short period binary (such as HW\,Vir); a $\gamma$\,Dor or $\delta$\,Sct star would be detected in the spectrum of J2344--3427, which was shown not to be the case by \citet{geier12}, thus we can exclude these A/F star pulsators from contaminating the light curve; and the lack of any IR excess in 2MASS photometry allows us to exclude sufficiently luminous stars which could have pulsations in our light curve. We are confident, therefore, that J2344--3427 is a single star which shows hybrid pulsations.

There are conflicting temperatures for J2344--3427 presented in the literature. \citet{mcdonald12} measured the effective temperature of J2344--3427 to be $37\,448\pm1\,648$\,K. This value is derived through SED fitting and does not take into account interstellar reddening. Furthermore, they use black-body fits which are not appropriate for hot stars. A more accurate result is presented by \citet{geier13b} who used their FEROS spectrum to derive a $T_{\rm eff}$ of $27\,000 \pm 500$\,K. Further to this, \citet{heber84} found a value of $28\,800 \pm 1\,500$\,K and a $\log g$ of $5.4\pm0.2$\,cm\,s$^{-2}$, and \citet{nemeth12} found a $T_{\rm eff}$ of $28\,390 \pm 265$\,K. These values from spectral fitting provide a much more reliable temperature estimate than that of \citet{mcdonald12}. Therefore we adopt $T_{\rm eff}=28\,000\pm250$\,K for this star by taking the weighted mean and its error from the spectroscopically derived temperatures. Such a temperature places the star in the hybrid star temperature range. Given that only one measurement of $\log g$ was found, we adopt that value.

Previous to {\it Kepler}, these hybrid pulsators were seen to have amplitudes much larger than their non-hybrid counterparts. However, \citet{reed10} have found pulsations in {\it Kepler} targets which would have been below ground-based detection limits. {\it Kepler} observations have also shown g\,modes to be of higher amplitude than the p\,modes in the same star, unlike previously observed hybrid sdBV stars, as well as J2344--3427 here.

The presence of both p and g\,modes in a single star is key to understanding the structure of the star as the different excitation mechanisms probe different depths. Further observations of J2344--3427 are required to confirm the presence of the g\,modes, and provide a full frequency solution for asteroseismological modelling. 

\section{Summary and conclusions}

As there are relatively few variable hot subdwarf stars known (Table\,\ref{tab:sdBV-cat}), the identification of further class members is key to understanding these stars. A previous attempt to identify new sdBV stars in the SuperWASP archive only resulted in the confirmation of previously known variables \citep{maxted08}. However, in this work, we have identified and analysed three new sdBV stars found in the SuperWASP archive. Two of these stars, J0902--0720 and J1938+5609, are of the short-period p-mode type, while the third, J2344--3427, is a hybrid pulsator. 

Amongst the short-period stars, J1938+5609 is at the lower temperature and surface gravity end of the distribution (Fig.\,\ref{fig:teff_logg}), and as such is mixed with the long-period pulsators in the $T_{\rm eff}-\log g$ plane. The reason for this is unclear.  However, the noise limits of the SuperWASP data do not allow us to rule-out the presence of low-amplitude g\,modes which would make this star a hybrid pulsator. It would then fall in the expected region in the $T_{\rm eff}-\log g$ plane. The position of J0902--0720 in the blue part of that plane suggests that g\,modes are unlikely to be found in this star.

J2344--3427 has been identified as a hybrid pulsator, and is found amongst the other hybrid stars in the $T_{\rm eff}-\log g$ plane. Given the location of J1938+5609 is similar to J2344--3427 in the $T_{\rm eff}-\log g$, we postulate that J1938+5609 star is also a hybrid pulsator. The amplitude of the g-mode pulsations in J2344--3427 are of order the noise level of J1938+5609 (in the same frequency range), further suggesting that J1938+5609 may be a hybrid pulsator (as defined in the limit of ground-based observations) with the signal lost in the noise. Additional observations, in the appropriate passband, are required to fully characterise all these targets. These observations may led to the identification of further, low-amplitude, pulsations not identified in our broadband observations.

The stars presented in this work demonstrate the ability of survey data to identify unknown sdBV stars, and thus expand the number which are available for in-depth analysis via targeted photometry and/or spectroscopy. Further to this, as ground-based surveys often target bright stars, the stars presented here are ideal candidates for follow-up observations by the TESS mission.

\section*{acknowledgements}
DLH thanks the STFC for financial support via grant ST/M000877/1. The SuperWASP project is funded and operated by Queen's University Belfast, the Universities of Keele, St. Andrews and Leicester, the Open University, the Isaac Newton Group, the Instituto de Astrofísica de Canarias, the South African Astronomical Observatory and by the STFC. This work was based on: service observations made with the WHT operated on the island of La Palma by the Isaac Newton Group in the Spanish Observatorio del Roque de los Muchachos of the Instituto de Astrofísica de Canarias; observations made with the Mercator Telescope, operated on the island of La Palma by the Flemmish Community, at the Spanish Observatorio del Roque de los Muchachos of the Instituto de Astrofísica de Canarias and; observations made with the Nordic Optical Telescope, operated by the Nordic Optical Telescope Scientific Association at the Observatorio del Roque de los Muchachos, La Palma, Spain, of the Instituto de Astrofisica de Canarias, with ALFOSC, which is provided by the Instituto de Astrofisica de Andalucia (IAA) under a joint agreement with the University of Copenhagen and NOTSA. We thank the referee, Simon Jeffery, for useful comments and suggestions which have improved the manuscript.

%%%%%%%%%%%%%%%%%%%%%%%%%%%%%%%%%%%%%%
\bibliography{sdBV-WASP-arXiv}

%%%%%%%%%%%%%%%%%%%%%%%%%%%%%%
\label{lastpage}

\end{document}